\documentclass{JHEP3} 

\usepackage{epsfig,multicol}

\newcommand\fverb{\setbox\pippobox=\hbox\bgroup\verb}
\newcommand\fverbdo{\egroup\medskip\noindent%
			\fbox{\unhbox\pippobox}\ }
\newcommand\fverbit{\egroup\item[\fbox{\unhbox\pippobox}]}
\newbox\pippobox
\newcommand{\beq}{\begin{equation}}
\newcommand{\eeq}{\end{equation}}
\newcommand{\ds}{{\tt DarkSUSY}}
\def\sss{\scriptscriptstyle}
\def\ie{{\em i.e.\ }}
\newcommand{\mpl}{M_{\rm\scriptscriptstyle Pl}}

\title{SUSY Dark Matter and Quintessence}

\author{Stefano Profumo and Piero Ullio\\
	SISSA/ISAS, via Beirut 2-4, 34013 Trieste, Italy\\
	E-mail: \email{profumo@sissa.it}, \email{ullio@sissa.it}}

\preprint{SISSA-79/2003/EP}

\abstract{We investigate the enhancement of neutralino relic density in the context of a realistic cosmological scenario with Quintessence. The accurate relic density computation we perform allows us to be sensitive to both cases with shifts in the abundance at the percent level, and to enhancements as large as $10^6$.
We thoroughly analyze the dependence on the supersymmetric spectrum and on the mass and composition of the lightest neutralino. We point out that supersymmetric models yielding a wino or higgsino-like lightest neutralino become cosmologically appealing in the presence of Quintessence.}


\begin{document} 

\section{Introduction}

The latest years have been marked by fast and steady progresses in
observational Cosmology. The picture emerging from the data is remarkably
self-consistent~\cite{Spergel:2003cb}, 
and, at the same time, it is significantly different from
that which used to be the standard lore less than a decade ago. 
The greatest surprise came with the discovery that the Universe
is accelerating~\cite{sne}, with most of its energy density currently in the
form of ``dark energy'', a component with negative pressure. 
In the simplest scenario compatible with these observations, 
with a cosmological constant term which has very recently
(on a cosmological timescale) matched and overtaken the rapidly 
decreasing matter term, a large degree of fine-tuning is needed in the
initial conditions in the early Universe. 
In recent years, alternative models have been proposed, in which dark energy is introduced
as ``Quintessence''~\cite{firstexp,firstquint,Steinhardt:nw,Ferreira:1997hj}, 
a dynamical term, such as a 
scalar field rolling down an effective potential, gravitationally 
coupled to the other components of the Universe and possibly tracking
the largest of them long before becoming the dominant one at the present epoch.
Besides driving the dynamics of the Universe today, such a term, which
has a time varying equation of state, may have played a r\^ole during
other stages in the evolution of the Universe: other issues in 
the ``canonical'' picture, not directly tested by observations so far,
might have been overlooked, a problem which has been addressed in 
a number of recent analysis and in different contexts, see, 
e.g.,~\cite{Salati:2002md,others}.

The latest Cosmological observations have confirmed that there is
a large discrepancy between the total amount of matter in the 
Universe and the baryonic component; in terms of the ratio $\Omega$
between the mean density and the critical density, the contribution of 
non-relativistic matter has been found~\cite{Spergel:2003cb} to lie 
in the range 
$\Omega_{m} h^2 = 0.135^{+ 0.009}_{-0.008}$
(here $h$ is the Hubble constant in units of 100 km s$^{-1}$ Mpc$^{-1}$;
$h = 0.71^{+0.04}_{-0.03}$~\cite{Spergel:2003cb}), much larger than 
that of the baryonic component
$\Omega_{b}h^2 = 0.0224 \pm 0.0009$. The nature of cold dark matter
remains one of the main puzzles in today's research. Probably,
the most natural approach to this problem is to suppose that
dark matter is just another of the thermal leftovers from the Early 
Universe. It is a remarkable coincidence that the thermal relic 
abundance of stable weakly-interacting massive
particles (WIMPs) is roughly of the order of $\Omega_{m}$: 
in several classes of extensions to the standard model (SM) of particle
physics such dark matter candidates arise naturally, the lightest
neutralino in supersymmetric extensions to the SM being the most
widely studied case.

As in any process involving departure from thermal equilibrium,
the freeze-out of WIMPs can be understood in terms of a particle 
interaction rate $\Gamma$ falling below the expansion rate of the 
Universe $H$. $\Gamma$ and its scaling with temperature are fully 
defined once the particle physics setup has been specified; what, instead, 
about $H$ and its thermal scaling? The freeze-out of WIMPs is 
predicted to take place at a temperature $T_{\sss\rm f.o.}$ in the range of
a few GeV or above,
much earlier, in the history of the Universe, than any of the processes 
which can be directly tested through cosmological observations. The 
standard prejudice is to extrapolate the snapshot of the Universe
we derive from tests of the standard big bang nucleosynthesis (BBN), 
at a temperature of about 1~MeV, backward to the WIMP freeze-out temperature, 
assuming a radiation dominated Universe. Implications of nonstandard cosmological scenarios at the time of decoupling were first envisaged and analyzed by Barrow~\cite{Barrow:ei} and by Kamionkowski and Turner~\cite{Kamionkowski:1990ni}: the various possibilities ranged from setups where a shear energy density component is generated by anisotropic expansion \cite{Barrow:ei,Kamionkowski:1990ni} to Brans-Dicke-Jordan cosmological models \cite{Kamionkowski:1990ni}. Following that line, in a recent analysis,
Salati~\cite{Salati:2002md} pointed out that predictions for the
WIMP relic abundance are significantly different if, on the other hand,
one supposes that the freeze-out happens during a period of
Quintessence domination, as it can be the case if the Quintessence
field undergoes a phase of ``kination'' (see the next Section for
a clarification on this point).

The underlining idea is rather simple: if on top of the radiation energy
density a Quintessence contribution is added, the Universe is forced
into having a faster expansion. A larger $H$ implies that the matching 
between annihilation rate and expansion rate takes place at a higher 
temperature, hence the WIMP equilibrium number density freezes
in at a larger value, giving a net increase in the final WIMP relic
abundance (even a modest increase in $T_{\sss\rm f.o.}$  induces large effects,
because the freeze-out of WIMPs happens when WIMPs are non-relativistic
and their equilibrium number density is along the exponentially 
suppressed Maxwell-Boltzmann
tail). We examine here this mechanism in detail: we implement 
a realistic prototype for Quintessence into the density evolution equation
describing the decoupling of a WIMP dark matter species, and
solve such equation numerically with the state of the art technique as 
developed in the  \ds\ 
package~\cite{Gondolo:2000ee,Gondolo:2002tz,Edsjo:2003us}. 
We are then in the position of
giving some general criteria to quantify the quintessential enhancement
of relic densities, and of illustrating what classes of models are more
sensitive to such an effect. We present results for a few sample cases
of neutralino dark matter candidates in the context of the minimal 
supersymmetric extension of the SM (MSSM). In particular we discuss
at length the case of higgsinos and winos, which are usually disregarded
as dark matter candidates, as their thermal relic abundance is typically exceedingly small in 
the standard cosmological scenario.

The outline of the paper is as follows: In the next section we introduce
our setup for quintessential Cosmology, and in section~3 we illustrate the 
possible r\^ole of Quintessence in shifting the neutralino relic abundance.
In section~4 we focus on the case of higgsinos and winos, while section~5
concludes.

\section{A r\^ole for Quintessence in the early Universe}\label{sec:quintsetup}

We consider a Friedmann-Robertson-Walker cosmology including, in addition 
to the usual radiation 
and matter terms, a Quintessence component, which we sketch as a single 
spatially-homogeneous scalar field $\phi$ with a potential 
$V\left(\phi\right)$. The energy density and pressure associated to this 
component are, respectively:
\beq
   \rho_{\phi}=\frac{1}{2} \left(\frac{d\phi}{dt}\right)^2 + 
   V\left(\phi\right)\;,\;\;\;\;\;\;\; 
   p_{\phi}=\frac{1}{2} \left(\frac{d\phi}{dt}\right)^2 - 
   V\left(\phi\right)\;.
\eeq
By formally writing the equation of state for $\phi$, 
$p_{\phi}= w \rho_{\phi}$, we see that $w$ can vary between $+1$ and
$-1$, going from the regime in which the kinetic term is much larger 
than the potential term (``kination'' phase~\cite{kina,Ferreira:1997hj}) 
to the opposite 
case when the field is frozen into one configuration and just behaves as
a cosmological constant term. 

To trace the Hubble expansion rate $H$ as a function of the temperature $T$, 
we need to solve the Friedmann equation (we assume a spatially flat Universe):
\beq
   H^2\left(T\right) = \frac{1}{3 \mpl^2} 
   \left[\rho_{r}\left(T\right) + \rho_m\left(T\right) + 
   \rho_{\phi}\left(T\right)\right]\;,
\label{eq:Hrate}
\eeq
coupled to the equation of motion for $\phi$:
\beq
   \frac{d^2\phi}{dt^2} + 3 H \frac{d\phi}{dt} + \frac{dV}{d\phi} = 0\;.
\label{eq:motion}
\eeq 
In Eq.~(\ref{eq:Hrate}) above, $\mpl \equiv 1/\sqrt{8 \pi G}$ is the 
reduced Plank mass, while $\rho_{r}$ and $\rho_m$ are, respectively, the energy 
density in radiation and in matter. To find $\rho_m(T)$ we
simply scale its current value with the appropriate equation of state;
for $\rho_{r}(T)$ the analogous procedure, often implemented in Quintessence
studies, is not accurate enough for our purpose. 
We write instead $\rho_r(T)$ as:
\beq
   \rho_{r}(T) = \frac{\pi^2}{30} g_{\rm eff}(T)\, T^4
\eeq
and compute  $g_{\rm eff}(T)$, the effective degrees of freedom coefficient, 
summing the contributions from all particles in the context at hand (\ie, 
in the supersymmetric extension to the SM we consider below,
we find $g_{\rm eff}(T)$ for each of the mass spectrum we generate,
including both standard model particles and the supersymmetric 
partners; details on how to compute $g_{\rm eff}(T)$ are given, e.g., in 
Ref.~\cite{gondologelmini}).

$\rho_\phi(T)$ and $H(T)$ can be derived once we specify
the initial conditions and the potential $V\left(\phi\right)$.
The set of initial conditions we resort to is not the most generic in a 
context of quintessential cosmology, as we wish to restrict to solutions 
in which  $\rho_\phi$  is initially larger than $\rho_{r}$, and then
is red-shifted away more rapidly than the radiation component
(this happens in the {\em kination} phase) so that the radiation 
domination epoch can 
start before the time of BBN. This corresponds, e.g., to the reheating 
scenario suggested in Ref.~\cite{spokoiny} and, more generically, 
in models that try to generate, through the same mechanism, the current 
inflationary period and a period of inflation in the 
early Universe (see, e.g., the quintessential inflationary models
of Ref.~\cite{infquint}). As regards the potential, we 
choose to work with the exponential form~\cite{firstexp}:
\beq
   V\left(\phi\right) = M_P^4 \exp\left(-\lambda \phi/M_P\right)\;,
   \label{eq:exppot}
\eeq
one of the simplest examples of potential which can lead to an attractor 
solution, self-tuning the contribution of $\phi$ to the energy density to 
the ``background'' contributions, \ie, in our case, radiation plus matter. 
We will select regions in the parameter space in which the attractor is
reached, so that we deal with a Quintessence model playing a main 
dynamical r\^ole in the recent past; actually the simple form for the
potential we picked is not going to give the behavior of $\rho_\phi$ 
observed today (the solution converges to the same equation of state 
as the largest background component), however slight modifications 
to it can introduce the right scaling, see, e.g., Ref.~\cite{Albrecht:1999rm,Sahni:1999qe}.
On the other hand, we checked in a few test cases that this slight
reshuffling of the potential can be introduced without varying 
significantly the dynamics of the field in the early Universe\footnote{We explicitly worked out the cases of the hyperbolic cosine \cite{Sahni:1999qe} and of the so-called AS potential, \ie an exponential potential with a power-low prefactor~\cite{Albrecht:1999rm}.};
we can therefore safely ignore details in the late time behavior of the field.
The exponential form itself is not critical for the results we will present, 
however, in order to have a viable prototype for Quintessence, 
one should restrict to setups which allow to reach tracking 
after going through a kination phase, a feature which is not shared by
all potentials which do admit an attractor solution; further examples of 
such setups are given in Ref.~\cite{rosati}.

\FIGURE[t]{\epsfig{file=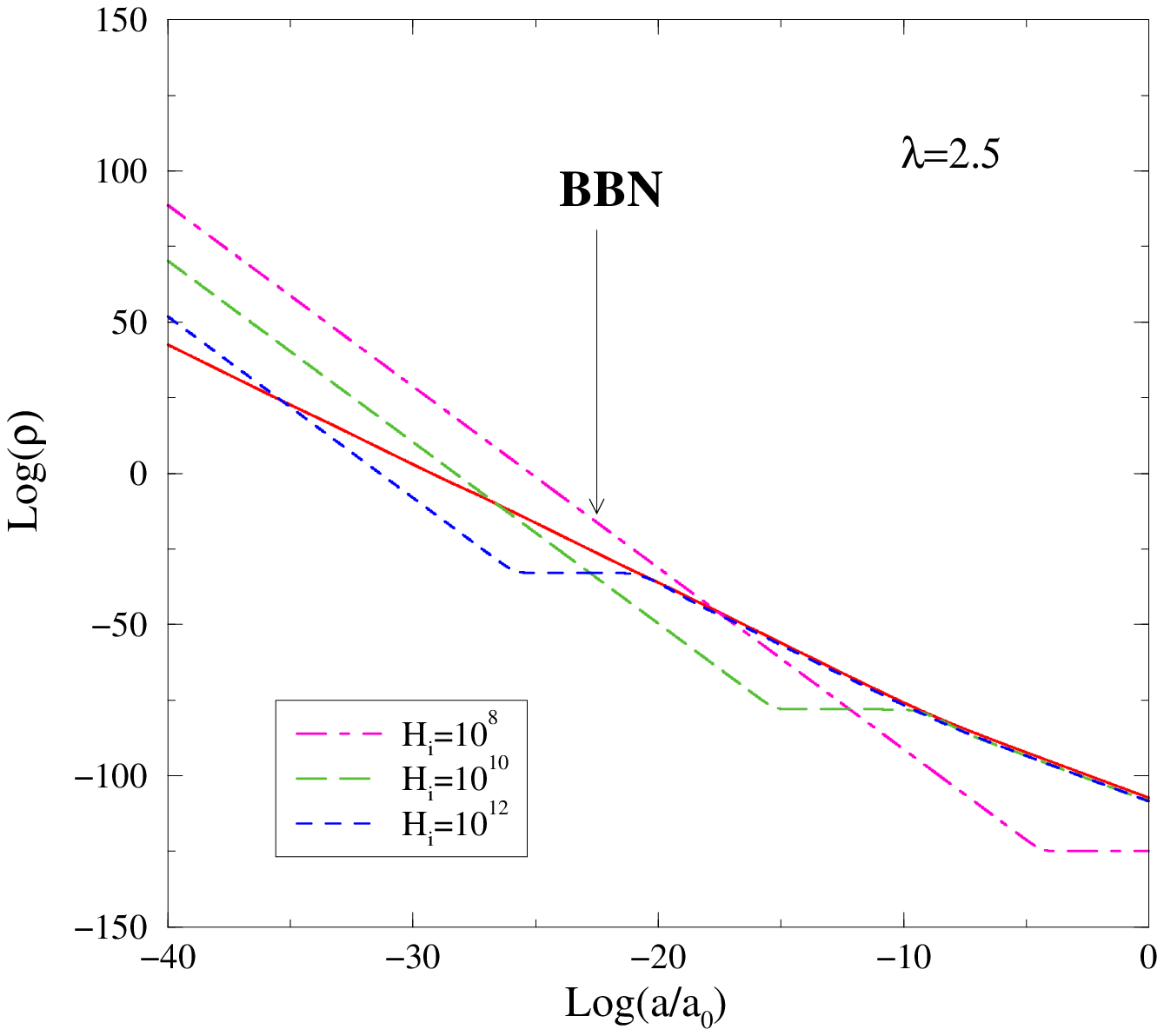,width=7.1cm}\qquad
\epsfig{file=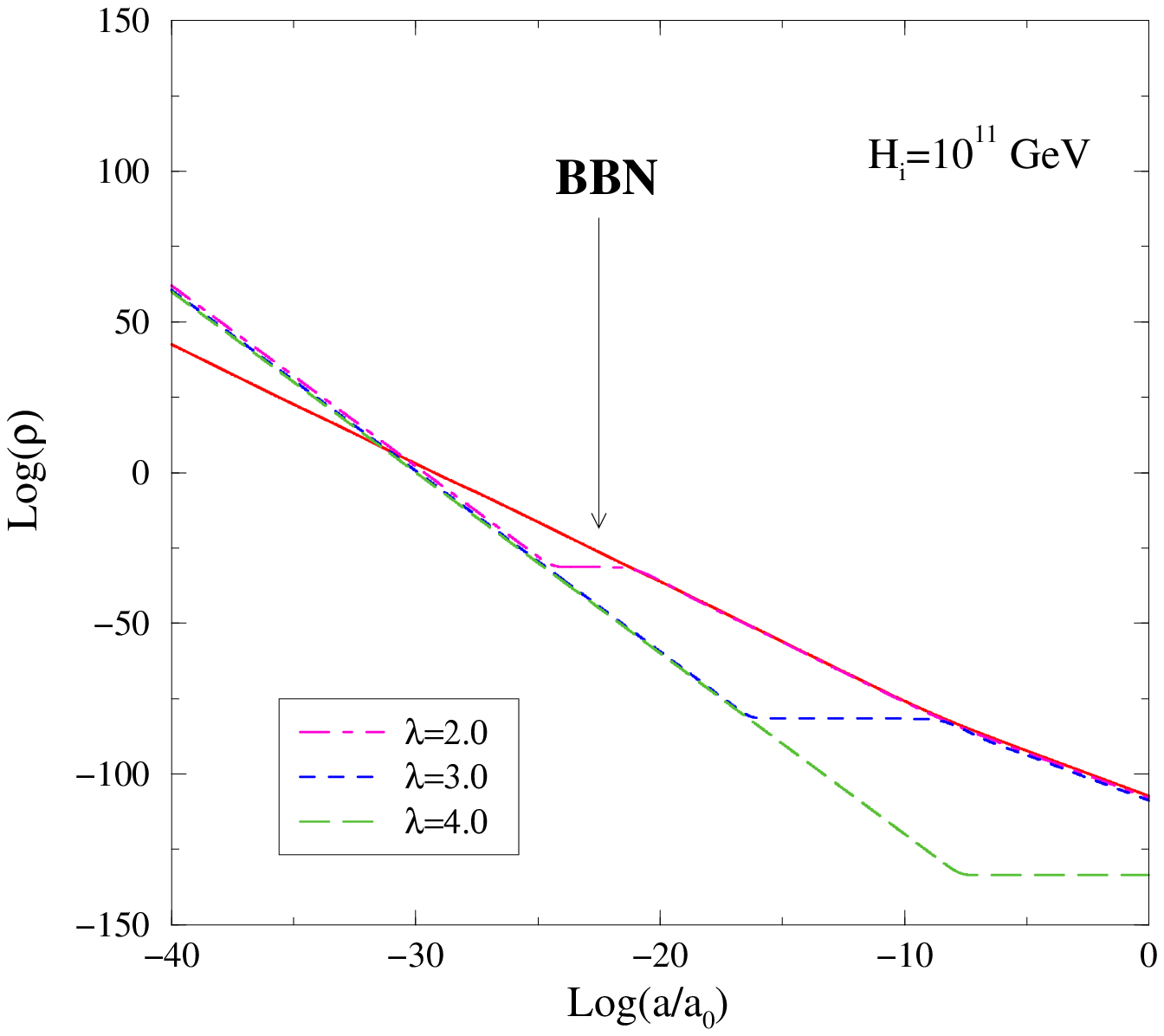,width=7.1cm} 
\caption{Sample cases of Cosmologies with a Quintessence field
in a kination phase dominating the total energy density in the 
early Universe. The solid lines refer to the sum of the 
radiation and matter energy density components, the other curves
to the Quintessence contributions. We have considered the case of
an exponential potential and show here the effect of
varying the two parameters in our setup: In the left panel three
sample values for the expansion rate at the end of inflation
$H_i$ are chosen, in the right panel the parameter $\lambda$ in
the exponential potential is varied.}
\label{fig:Hilambda}}


The possibility of having a kination regime with
the exponential potential introduced above was examined in detail 
by Ferreira and Joyce in Ref.~\cite{Ferreira:1997hj}; we closely follow 
their analysis here, choosing, in particular, the same set of initial
conditions. In terms of $H_i$, the expansion rate at the end of
inflation, we fix:
\beq
   V\left(\phi_{\rm init}\right) = 3 M_P^2 H_i^2\;, \quad
   \left.\frac{d\phi}{dt}\right|_{\phi_{\rm init}} = 0\;, \quad
   {\rm and} \quad
   \rho_r\left(T_{\rm init}\right) = 10^{-3} H_i^4\;; 
\eeq
the energy scale for $H_i$ can be in the range 
$[10^8$,$10^{16}]$~GeV~\cite{Ferreira:1997hj}\footnote{From WMAP data a more stringent, though model dependent, upper bound has been derived \cite{Peiris:2003ff}: \mbox{$H_i\lesssim 3\times 10^{14}{\rm GeV}$}, possibly cutting out a slice in the parameter space. This, however, does not affect the regime of interest in the present paper, which lies in the {\em low} $H_i$ range.}.
For any given choice of the two parameters in our model, \ie
the parameter $\lambda$ in the exponential potential and $H_i$, 
the system of differential equations (\ref{eq:Hrate}) and 
(\ref{eq:motion}) can be easily solved numerically. A few sample 
cases (see also Ref.~\cite{Ferreira:1997hj}) are shown in 
Fig~\ref{fig:Hilambda} where we plot energy density components as 
a function of the Universe scale factor $a$. Solid lines refer to
the sum of the matter and radiation component (with an ankle at 
the value when the former, scaling as $a^{-3}$, takes over the 
latter, which scales instead as $a^{-4}$); Quintessence energy density 
components (scaling as $a^{-6}$ in the initial kination phase) are also 
shown. An increase in $H_i$ shifts the initial conditions to a higher
temperature and, at the same time, lowers the initial ratio
$\rho_\phi / \rho_{r}\propto H_i^{-2}$:
the effect is
obviously to anticipate the radiation dominated phase and
the tracking phase. An increase in $\lambda$ makes the potential
steeper, henceforth the kination phase to last longer; from the point of 
view of the early time behavior, there is just a slight decrease of the 
normalization of $\rho_\phi$ in the kination phase due to a quicker transition from the initial condition (with $\phi$ at rest) 
into the fast rolling phase 
(for this specific potential the normalization of $\rho_\phi$ at 
tracking scales instead as $1/\lambda^2$~\cite{Ferreira:1997hj}, but, as 
already mentioned, we can neglect here details in the late time behavior 
of $\rho_\phi$).

In Fig~\ref{fig:allowed} we show the region of the parameter space 
$H_i$ versus $\lambda$ which is compatible with the limits from BBN. The latter are often reported in terms of bounds on extra relativistic degrees of freedom $\Delta N_{\rm eff}$ at BBN, which may be translated in an upper bound on the fraction $\Omega_\phi\equiv\rho_\phi/\rho_r\lesssim0.1$ for $\Delta N_{\rm eff}\simeq1.0$.
In the dark shaded region the Quintessence field produces a 
contribution to the energy density at 1~MeV which exceeds the stated constraint. 
In the lower part of the figure this happens because the $\phi$ field is still 
in the kination phase, and exceeds at the time of BBN the bound $\Omega_\phi\lesssim0.1$; in the upper-left part, instead, the field
is already in the attractor solution at 1~MeV, and again violates the BBN bound. The light shaded region
indicates, on the other hand, the cases in which the ratio between $\rho_\phi$
and the background components gets overly damped, so that tracking does 
not take place before the present epoch ({\em overshooting solutions}); both this cases are disregarded in the present work.

\FIGURE[t]{\epsfig{file=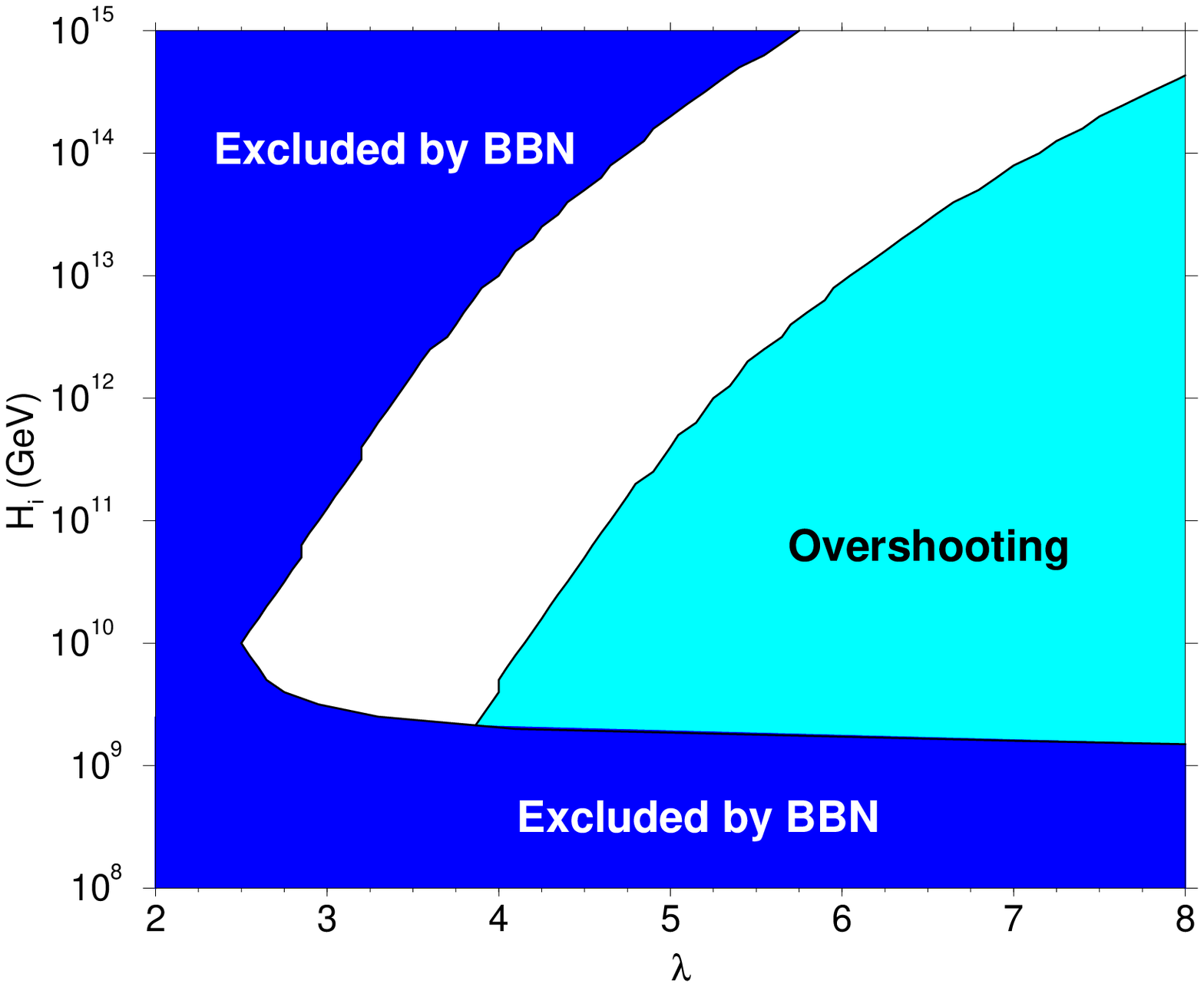,width=9cm}
\caption{Region of the parameter space $H_i$ versus $\lambda$ which is 
excluded by the limits from standard big bang nucleosynthesis. Also
indicated is the region where the ratio Quintessence energy density
to background energy density gets exceedingly small and tracking does 
not take place before the present epoch.}
\label{fig:allowed}
}


\section{Quintessence and neutralino dark matter}\label{sec:quitneutralino}

The density evolution equation describing the decoupling of a species
from the thermal bath is the Boltzmann equation:
\begin{equation} \label{eq:Boltzmann2}
  \frac{dn}{dt} =
  -3Hn - \langle \sigma_{\rm{eff}} v \rangle 
  \left( n^2 - n_{\rm{eq}}^2 \right)\ ,
\end{equation}
where $n$ is the particle number density, $n_{\rm{eq}}$ the corresponding equilibrium
density, and $\langle \sigma_{\rm eff} v \rangle$  the 
thermally-averaged pair annihilation cross section.
 In case {\em coannihilation effects} \cite{Binetruy:1983jf,Griest:1990kh,Edsjo:1997bg} are present, \ie when extra particles are nearly degenerate in mass
with the (lightest) stable one, and share a quantum number with it, $n$ is defined as the total number density summed over all coannihilating 
particles, and $\langle \sigma_{\rm eff} v \rangle$ contains a weighed sum
over all relevant rates, see, e.g.,~\cite{Edsjo:1997bg} for details. 
Rephrasing the density 
evolution equation (\ref{eq:Boltzmann2}) in terms of $Y=n/s$, 
where $s$ is the entropy density, as a function of the ratio between
the particle mass and the temperature, $x = m/T$, one gets the equation
\begin{equation} \label{eq:Boltzmann3}
  \frac{dY}{dx} = \frac{\langle \sigma_{\rm{eff}} v \rangle}{3H} \frac{ds}{dx}
  \left( Y^2 - Y_{\rm{eq}}^2 \right).
\end{equation} 
This equation can be recast as
\begin{equation} \label{eq:Boltzmann4}
 \frac{x}{Y_{\rm eq}} \frac{dY}{dx} = \frac{\Gamma}{H}\, \frac{x}{3s}\frac{ds}{dx}\left(\frac{Y^2}{Y^2_{\rm eq}}-1\right) \simeq -\frac{\Gamma}{H}\, \left(\frac{Y^2}{Y^2_{\rm eq}}-1\right),
\end{equation}
where the r\^oles of the annihilation rate $\Gamma\equiv n_{\rm eq}\langle \sigma_{\rm eff} v \rangle$ and of the expansion rate $H$ become more
transparent: When $\Gamma/H\gg1$, thermal equilibrium holds, that is $Y\simeq Y_{\rm eq}$, while in the opposite regime $Y_{\rm eq}\ll Y$. As a rule-of-thumb, the freeze-out occurs when the ratio $\Gamma/H\sim1$. An increase in $\langle \sigma_{\rm eff} v \rangle$ yields a decrease in the freeze-out temperature. On the other hand, as already mentioned, if we increase $H$ adding an extra component on top of the radiation
component, such as in the quintessential scenario outlined in the
previous section,  the freeze-out takes place at larger temperatures, and the
thermal relic density can be sharply {\em enhanced}. Even a modest
shift in $T_{\sss\rm f.o.}$ has a dramatic effect, with 
an increase in the relic density of typical dark matter candidates
which can be up to several orders of magnitude: This is due to the fact 
that the thermal 
equilibrium number density scales {\em exponentially} with the temperature,
\ie $n_{\rm{eq}} \propto \left(m T\right)^{3/2} \exp(-m/T)$.

To better quantify the enhancement, rather than introducing a simplified 
toy-model and deriving an approximate solution to Boltzmann equation as 
done in Ref.~\cite{Salati:2002md}, we interface the expansion rate of the
Universe $H$ we have derived in the Quintessence scheme described in 
sec.~2 into the accurate numerical solution of Boltzmann equation
included in the \ds\ software 
package~\cite{Gondolo:2002tz,Gondolo:2000ee,Edsjo:2003us}.
For any given set of initial conditions in the Early Universe, and 
for any particle physics setup with given mass spectrum and particle
couplings, we then have  a tool to compute WIMP relic abundances
with an accuracy of the order of 1\%, \ie at the precision level of
upcoming measurements of $\Omega_m$.

\subsection{The particle physics setup}

In our numerical study we focus on neutralino dark matter candidates
within the context of the MSSM. We recall that neutralinos are mass 
eigenstates derived from the linear superposition of neutral gauginos,
the bino $\widetilde{B}$ and the wino $\widetilde{W}^3$,
and higgsinos: 
\begin{equation}
   \widetilde\chi_i=N_{i1}\widetilde{B}+N_{i2}\widetilde{W}^3+N_{i3}
   \widetilde{H}_1^0
   +N_{i4}\widetilde{H}_2^0\;.
\end{equation}
The lightest of them is often also the lightest supersymmetric particle 
(LSP), and, thanks to $R$-parity, stable; such neutral and stable particle is potentially 
a good WIMP dark matter candidate. We define 
{\em wino fraction} of the lightest neutralino the quantity $|N_{12}|$ 
and {\em higgsino fraction} the quantity $\sqrt{|N_{13}|^2+|N_{14}|^2}$. 
If the lightest neutralino is wino-like, \ie\ if it has a dominant wino 
fraction, it is also approximately degenerate in mass with the lightest 
chargino; analogously, if it is higgsino-like, there will be a 
quasi-degeneracy both with the next-to-lightest neutralino and with the 
lightest chargino. These degeneracies yield large coannihilation 
effects. Particular models which entail a wino or higgsino-like LSP will 
be discussed in sec.~\ref{sec:winohiggsino}.

As shown in eq.~(\ref{eq:Boltzmann3}),
the neutralino relic density strongly depends on the neutralino-neutralino 
annihilation rate. It is well known that, in the case of bino-like LSPs, 
this rate tends, in general, to be not large enough to avoid an overproduction
of thermal relic neutralinos. It is often quoted that, within the 
so-called constrained MSSM, only four regions of the parameter space give 
rise to relic densities compatible with the current cosmological upper
bound on $\Omega_m$:
(1) A {\em bulk} region, with a light neutralino and, typically, tight 
accelerator constraints ({\em e.g.}, $BR(b\rightarrow s\gamma)$, 
$\delta a_\mu$ or the bound on the mass of the lightest $CP$-even Higgs boson); 
(2) A {\em coannihilation} region with the NLSP, which in most of the 
parameter space is the lightest stau; 
(3) A {\em funnel} region, where the neutralino mass is close to {\em half} 
the mass of a neutral Higgs boson, most often the
$CP$-odd Higgs boson $A$, entailing strong {\em resonance} 
effects which drastically enhance the neutralino-neutralino 
cross section; 
(4) A {\em focus point} region, at large $m_0$, where electroweak symmetry 
breaking conditions dictate a sizeable higgsino fraction in the LSP composition.

In a more general setting, when one relaxes some assumptions of the 
constrained MSSM in the context of supergravity, or resorts to other 
phenomenological supersymmetry breaking mechanisms, the lightest neutralino 
may naturally feature large higgsino or wino components. 
For instance, if the gauge kinetic function depends on a non-singlet chiral 
superfield, whose auxiliary $F$ component is responsible for the gaugino 
mass generation, particular (fixed) ratios exist between the three gaugino 
masses at the GUT scale \cite{Ellis:1985jn,Drees:1985bx}. In the simplest 
case of $SU(5)$ GUTs, the three non-singlet representations are the {\bf 24}, 
the {\bf 75} and the {\bf 200}. The latter two (largest) representations 
yield {\em higgsino}-like LSPs. In what follows, we will pick as a sample case the 
{\bf 200}, which has the virtue of allowing a larger parameter space, 
particularly at large $\tan\beta$, with respect to the {\bf 75}.

\subsection{Relic abundance enhancement}

\FIGURE[t]{\epsfig{file=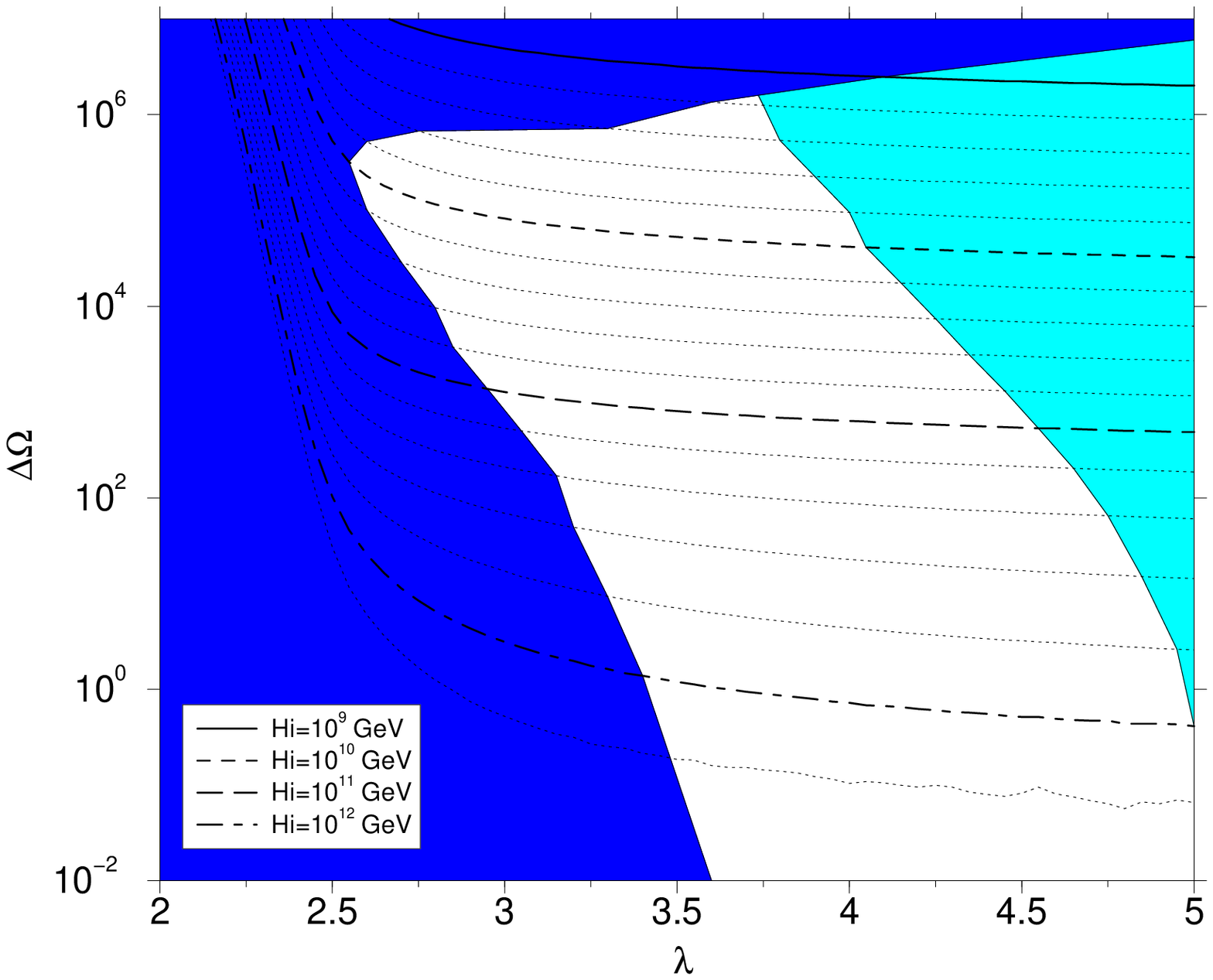,width=9cm}
\caption{Allowed region and iso-$H_i$ curves in the ($\lambda,\ \Delta\Omega$) plane. The dark blue region does not fulfill BBN constraints, while in the light blue zone no tracking takes place before nowadays cosmological times ({\em overshooting}). We show with bold lines the curves with the same $H_i=,\ 10^9,\ 10^{10},\ 10^{11}$ and $10^{12}$. Also shown, in dotted lines, curves at equal {\em logarithmic spacing} from those in bold lines, \ie\ at around 1.58, 2.51, 3.98 and 6.31 times the upper bold line $H_i$.}\label{fig:isohi}
}


We choose as indicator of the enhancement produced by a quintessential 
cosmology onto the neutralino relic density the quantity
\begin{equation}\label{eq:deltaomega}
\Delta\Omega\equiv\frac{\Omega_Q-\Omega}{\Omega},
\end{equation}
$\Omega h^2$ being the neutralino relic density in a cosmology {\em without} 
Quintessence (standard case), and $\Omega_Q h^2$ that {\em with} Quintessence. 
In Fig~\ref{fig:isohi} we show a first example of how large an enhancement 
we can obtain in our Quintessence setup. In the ($\lambda$,\ $\Delta\Omega$) 
plane, we plot curves at fixed Hubble rate at the end of inflation $H_i$; 
in this figure we have focussed on a specific neutralino candidate, selecting the point in the {\em bulk} region 
of minimal supergravity defined by the supersymmetric parameters 
$M_{1/2}=2300\ {\rm GeV},\ m_0=3000\ {\rm GeV},\ \tan\beta=45,\ A_0=0$ 
and $\mu>0$. With such choice we avoid, for the moment, coannihilation 
 or resonance effects, and consider the case of a heavy (bino-like) 
neutralino ($m_\chi\approx \ {\rm 1\ TeV}$), for which the enhancement 
effects are maximized: as we will 
discuss in detail below, the heavier the freezing-out particle, the larger 
the quintessential effects at decoupling, and therefore the
larger $\Delta\Omega$. The shadings in the figure reproduce those of 
Fig~\ref{fig:allowed}: dark blue corresponds to a scenario ruled out by 
BBN constraints, while light blue indicates overshooting (\ie\ tracking is
 not achieved at current cosmological times). The largest enhancement allowed
is of the order  $10^6$ and it is about the largest enhancement we can get
with a WIMP of mass smaller than 1 TeV. 
Moreover, we see that the minimum variation
we are sensitive to, an enhancement of the order of 1\% occurs for
$H_i$ slightly above $10^{12}$: This is the limit around which, in 
the process of decoupling of a WIMP with mass below  1 TeV, we are not
sensitive any longer to the presence of a kination phase. 

\FIGURE[t]{\epsfig{file=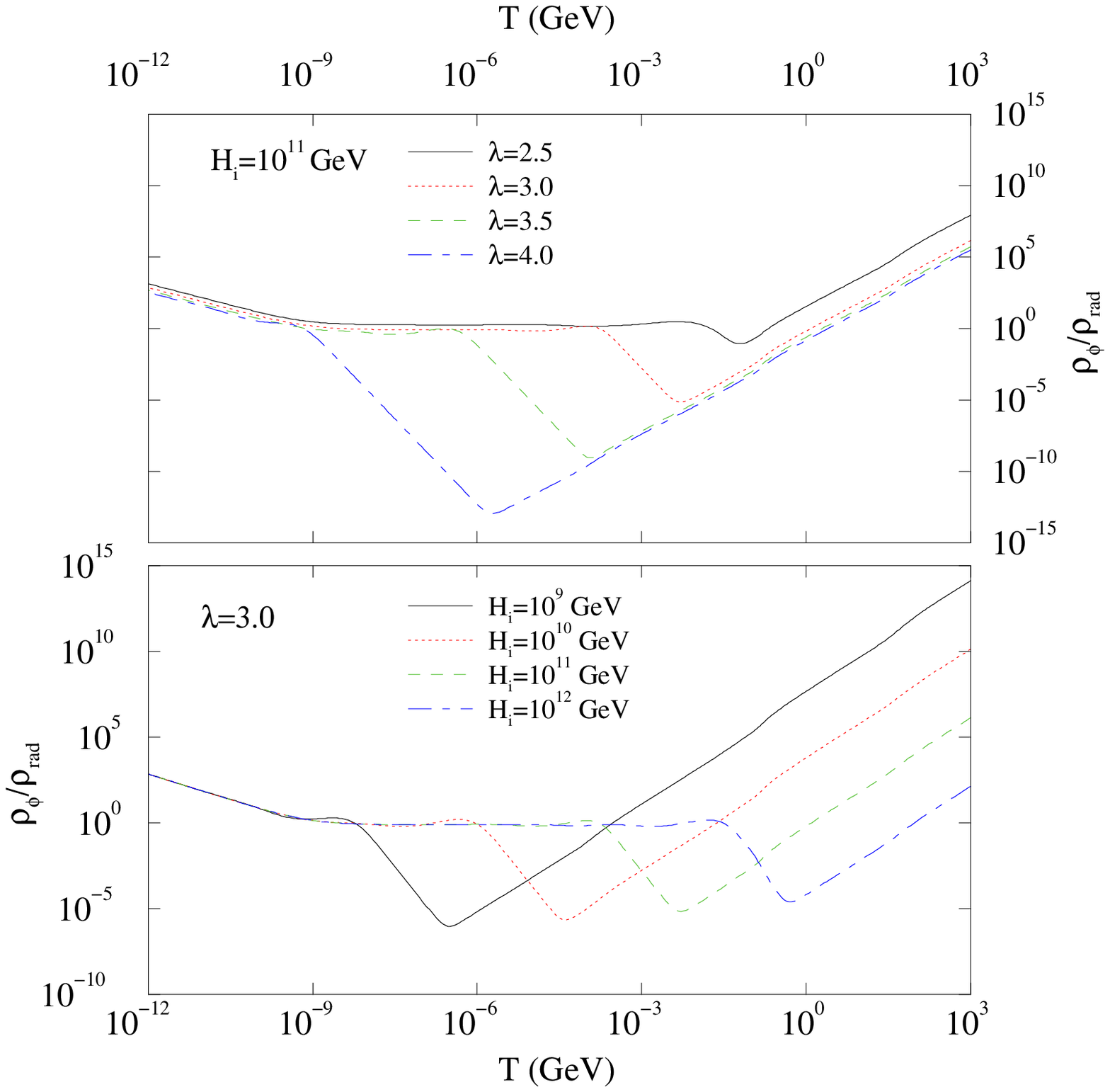,width=9cm}
\caption{Ratio Quintessence energy density
to radiation energy density as a function of temperature, for a
few sample choices of the paramters $\lambda$ and $H_i$.
}
\label{fig:ratiophirad}
}


As a last remark on Fig~\ref{fig:isohi},
we comment on the shape of the iso-$H_i$ curves at low 
$\lambda$: the abrupt raise in 
$\Delta\Omega$ is due to the fact that tracking takes place, at low 
$\lambda$, as soon as $\rho_\phi\simeq\rho_{r}$, without an 
overshooting period where  $\rho_\phi\ll\rho_{r}$. This of course 
yields a large quintessential contribution to $H$ at all time scales 
relevant to neutralino freeze-out, contrary to the case where an 
overshooting ($\rho_\phi\ll\rho_{r}$) takes place. However,
these sudden rises in $\Delta\Omega$ happen in the region excluded
by BBN constraints, as the tracking phase cannot appear before BBN.
Rejecting these cases,  we find, as expected, that all 
large enhacements take place for models in which the field $\phi$ is 
in a kination phase, $\rho_\phi\propto a^{-6}$. It appears then
natural, rather than searching for a parametrization of $\Delta\Omega$ 
as a function of $\lambda$ and $H_i$, to introduce, as one single relevant
parameter, the relative normalization of $\rho_\phi$ and 
$\rho_{r}$ at a given temperature. We do not have the freedom
to choose such temperature as low as, say, the BBN scale $T\simeq1$~MeV, 
as it is not unusual for the field $\phi$ to be, at this temperature, 
in its transient phase between kination and tracking (this is true 
even for our Quintessence model in which the transient is always
very sharp, with $\phi$ behaving like a cosmological constant; 
more generic models allow for softer and more complicated transients).
This is implicit in Fig~\ref{fig:Hilambda} and  more
directly illustrated in Fig~\ref{fig:ratiophirad}, where the ratio
$\rho_\phi/\rho_{r}$ is plotted versus the temperature for a few
sample choices of the parameters $\lambda$ and $H_i$: in the kination
phase this ratio scales nearly as $T^2$ (behaviour of all curves
for the largest temperatures display), but this is not necesseraly true 
at 1~MeV.

In order to quantify the amount of Quintessence relevant for the neutralino 
relic density enhancement, we choose to resort to the parameter:
\beq
\xi_\phi\equiv
\frac{\rho_\phi}{\rho_{r}}\left(T^{\rm NQ}_{\sss\rm f.o.}\right),
\label{eq:xi}
\eeq
where we defined $T^{\rm NQ}_{\rm\sss f.o.}$ the
temperature of neutralino freeze-out in absence of Quintessence 
(we use here the convention that the freeze-out temperature is the
temperature at which the abundance of the relic species is 50\%
larger than the equilibrium value, as computed from the full
numerical solution of the density evolution equation).
Avoiding the reference to a single cosmological temperature, the 
parameter $\xi_\phi$ entails a nearly model-independent estimate of 
the relevant amount of Quintessence. In order to have a rough estimate of this effect, we can consider as a first order approximation of $\Omega$ the following expression:
\begin{equation}
\Omega=\frac{\rho_{\chi}(T_0)}{\rho_c}=\frac{m_{\chi}n(T_0)}{\rho_c}\simeq\frac{m_{\chi}n_{\rm eq}(T_{\sss\rm f.o.})\ s(T_0)}{\rho_c\ s(T_{\sss\rm f.o.})}\simeq\frac{m_\chi\ s(T_0)}{\rho_c\ s(T_{\sss\rm f.o.})}\frac{H(T_{\sss\rm f.o.})}{\langle\sigma_{\rm eff}v\rangle},
\end{equation}
where in the third step we introduced the approximation that $Y(T_0)\simeq Y_{\rm eq}(T_{\sss\rm f.o.})$, while in the last step we assumed that the freeze-out temperature is {\em defined} by the relation $\Gamma(T_{\sss\rm f.o.})=H(T_{\sss\rm f.o.})$.

In the context of a quintessential cosmology, the expansion rate gets an additional factor given by:
\begin{equation}\label{eq:quintfactor}
H^{\rm NQ}(T^{\rm NQ}_{\sss\rm f.o.})\rightarrow H^{\rm NQ}(T^{\rm Q}_{\sss\rm f.o.})\sqrt{1+\frac{\rho_\phi}{\rho_{r}}(T^{\rm Q}_{\sss\rm f.o.})}\simeq H^{\rm NQ}(T^{\rm Q}_{\sss\rm f.o.})\sqrt{1+\xi_\phi\frac{(T^{\rm Q}_{\sss\rm f.o.})^2}{(T^{\rm NQ}_{\sss\rm f.o.})^2}},
\end{equation}
where the last approximate equality holds under the simplifying hypothesis that $\rho_\phi\propto a^{-6}$.
Since the actual shift in the freeze-out temperature should not be dramatic, \ie $T^{\rm Q}_{\sss\rm f.o.}\approx T^{\rm NQ}_{\sss\rm f.o.}$, we expect the relic abundance enhancement to be roughly driven by the factor $\sqrt{1+\xi_\phi}$.

\subsection{Wandering about the SUSY parameter space}
\label{sec:susyparameterspace}

As already alluded, the phenomenon of  quintessential enhancement 
of the neutralino relic density should depend not only on the particular 
quintessential setup, but also on the details of the choice of the
supersymmetric dark matter candidate. In particular, we will show here 
that there is a dependence (1) on the lightest neutralino mass, since this 
sets the overall energy scale at which neutralinos freeze-out, (2) on the 
details of the composition of the lightest neutralino in terms of its bino, 
wino and higgsino fractions, and (3) on the supersymmetric mass spectrum, 
especially if it entails coannihilation processes or resonances. 

\FIGURE[t]{\epsfig{file=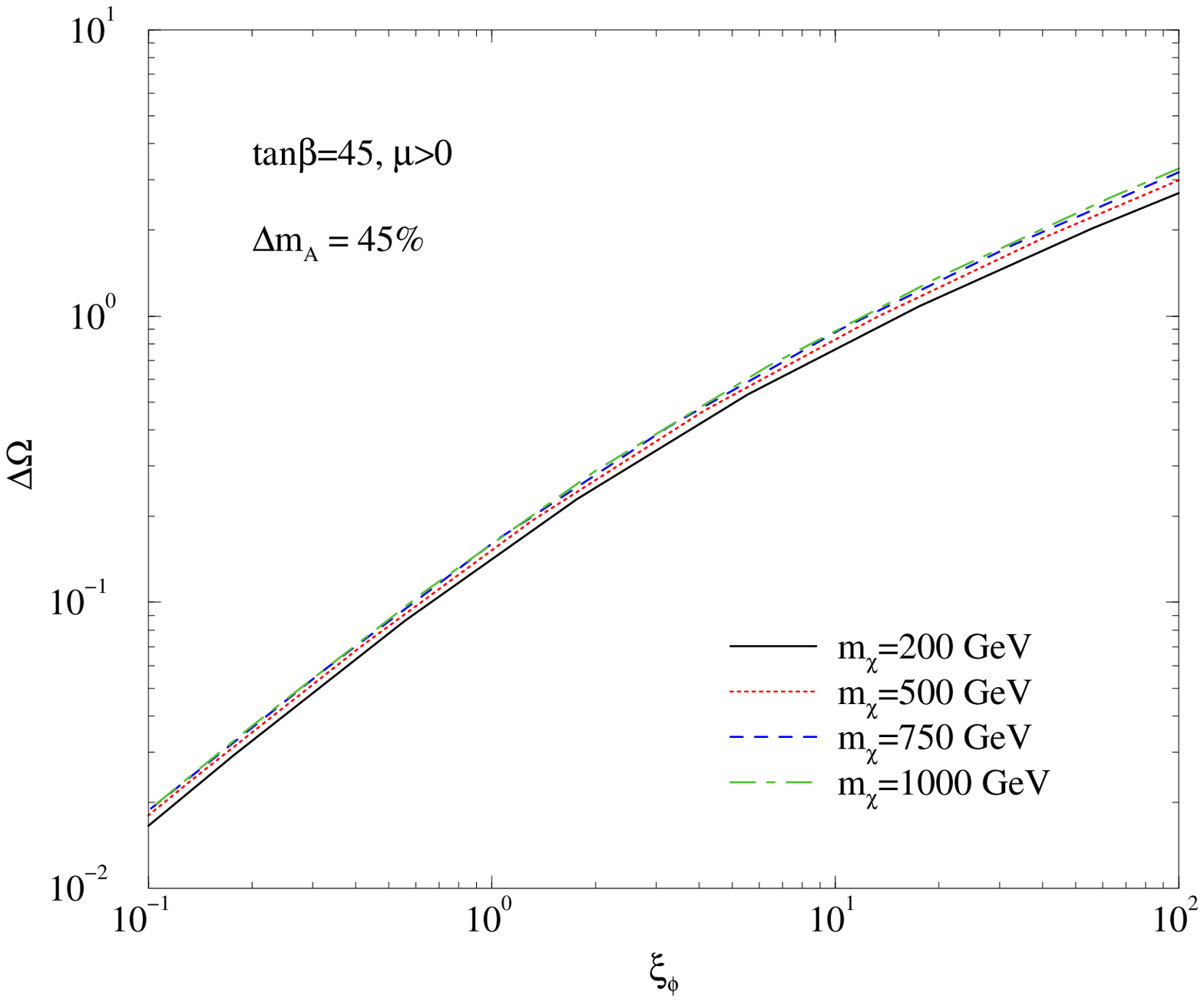,width=7.1cm}\qquad
\epsfig{file=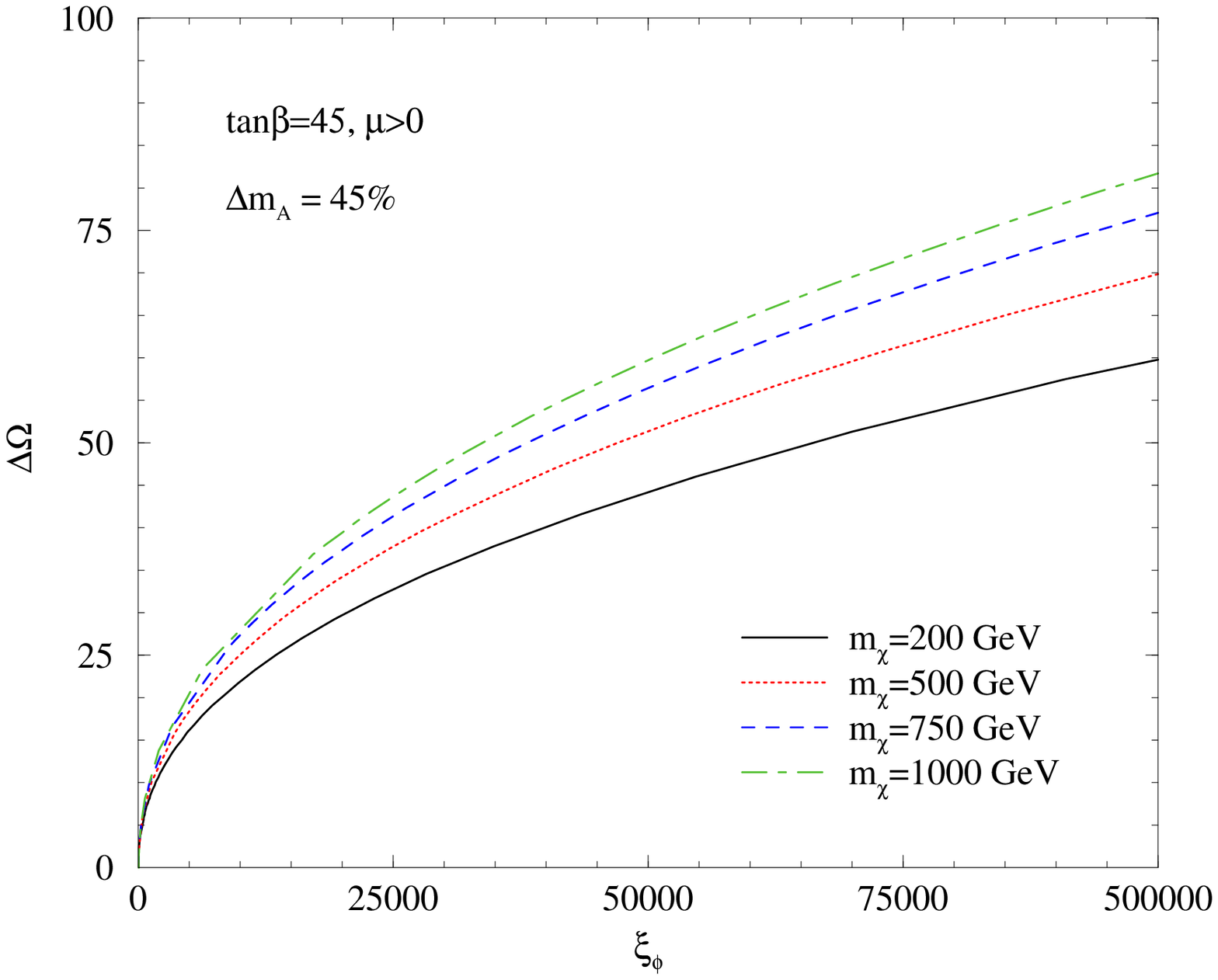,width=7.1cm}
\caption{The relic density enhancement $\Delta\Omega$ as a function of the Quintessence-to-radiation ratio at neutralino freeze-out $\xi_\phi$ for four SUSY models with neutralino masses equal to 200, 500, 750 and 1000 GeV. In all cases, $\tan\beta=45$, $\mu>0$ and $A_0=0$. The parameters $m_0$ and $M_{1/2}$ are tuned in order to obtain the desired neutralino mass and a mass splitting $\Delta m_A=(2m_\chi-m_A)/(2m_\chi)=45\%$.}\label{fig:enmass}
}


We focus first on the r\^ole of the neutralino mass. 
In the left panel of Fig~\ref{fig:enmass} we plot $\Delta\Omega$ versus
$\xi_\phi$, the ratio of the Quintessence energy density to the radiation 
energy density computed at the freeze-out temperature, for four neutralinos
in the bulk region of the minimal supergravity model and with
masses $m_\chi=200,\ 500,\ 750$ and 1000~GeV. The supersymmetric parameters 
have been chosen in order to exclude coannihilations or resonance effects, 
and the neutralinos are always almost completely bino-like. As regards the cosmological setup, we fixed $\lambda=3.5$, and let $H_i$ vary in order to span over $\xi_\phi$. As already 
mentioned, larger masses yield larger enhancements for a given
Quintessence setup; we can deduce however from the figure that, to a good 
approximation, for relatively low quintessential fractions, the quantity 
$\xi_\phi$ fixes $\Delta\Omega$. In particular we find that a 10\% 
enhancement in the relic abundance is obtained if at 
$T^{\rm NQ}_{\rm\sss f.o.}$ the Quintessence energy density is about 
60\% of the radiation energy density, and that a 100\% effect is generated
by $\xi_\phi \simeq 12$. As the amount of Quintessence grows, the spread
in enhancements for different masses, and at a fixed $\xi_\phi$, starts 
to increase, as shown in the right panel of Fig~\ref{fig:enmass}; 
$\Delta\Omega$ is in the range 60 to 80 for the largest value of 
the ratio we display, $\xi_\phi = 10^5$, a 15\% spread around the mean 
value: this reminds us that the freeze-out is not a sudden process
characterized by a single energy scale, and if an accurate estimate
of relic abundance is needed, a careful numerical solution to
the Boltzmann equation has to be implemented.

\FIGURE[t]{\epsfig{file=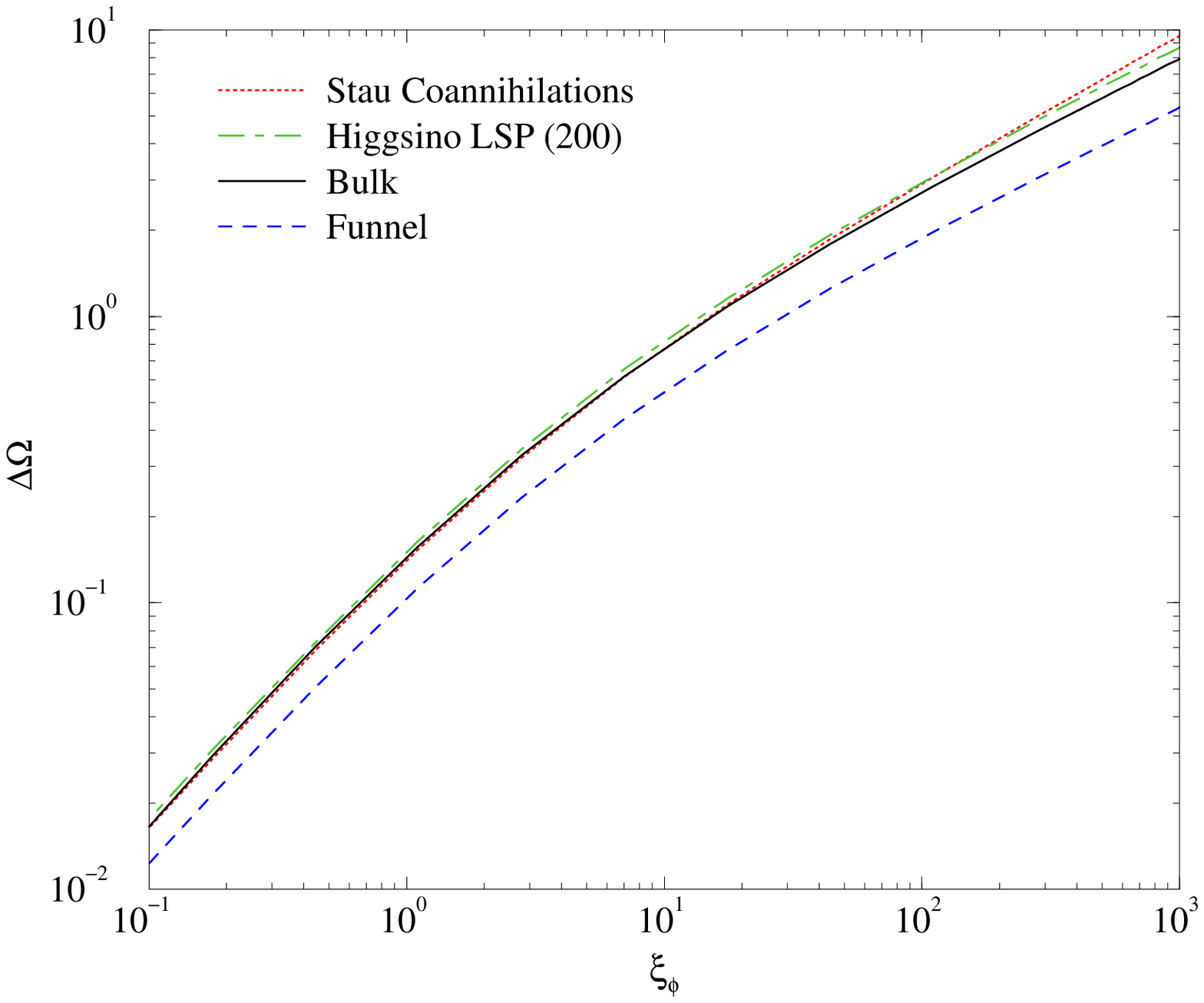,width=7.1cm}\qquad
\epsfig{file=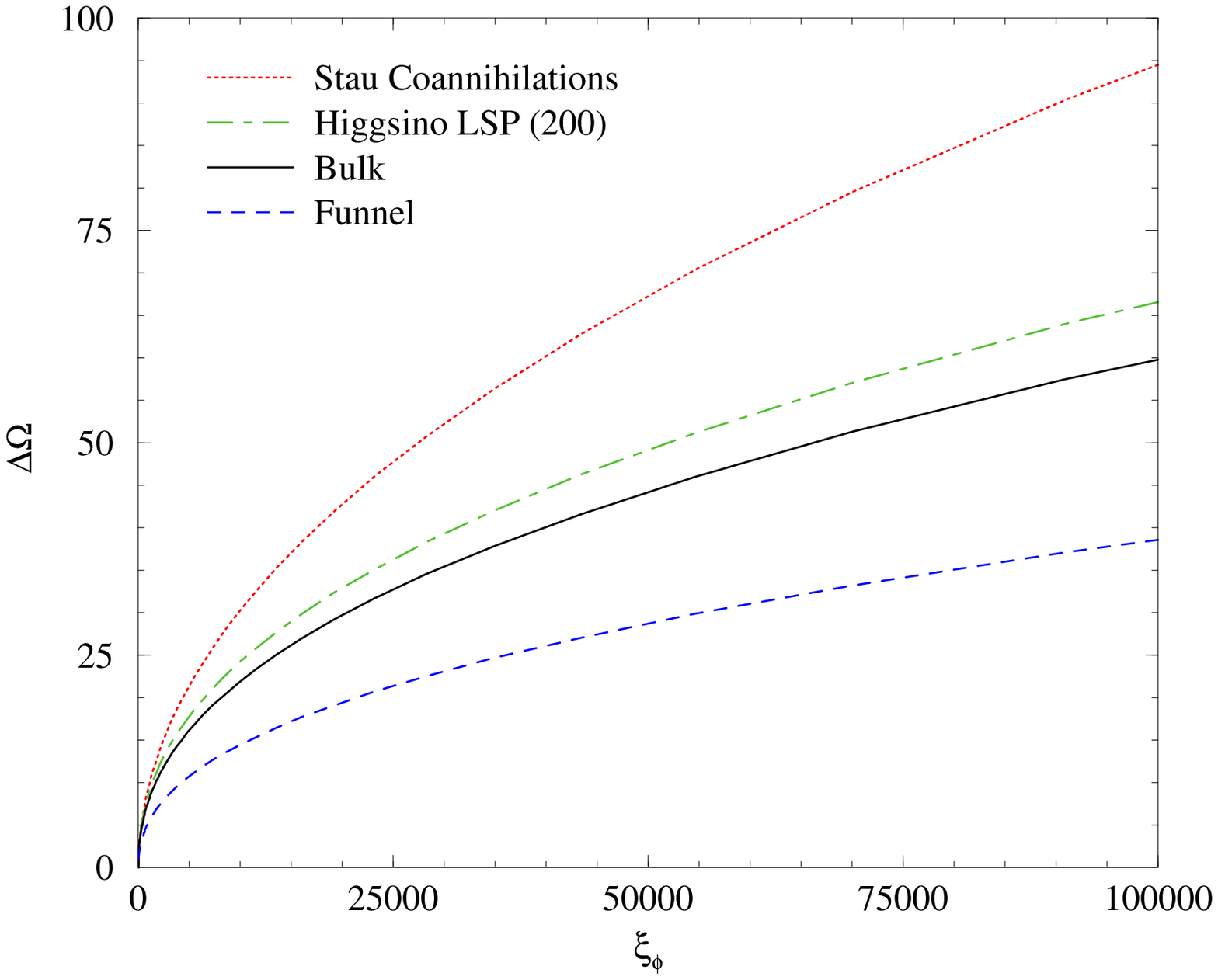,width=7.1cm}
\caption{The quintessential enhancement of the neutralino relic density for the four {\em benchmark} models of Tab.\protect{\ref{tab:benchmark}} 
as a function of $\xi_\phi$, respectively at large $\xi_\phi$ in a linear scale ({\em left})and at small $\xi_\phi$ in a logarithmic scale ({\em right}).}\label{fig:casestudyCOMP}
}


In order to study the dependence of the quintessential enhancement on the particular structure of the lightest neutralino, we focus on four {\em benchmark points}, chosen {\em ad hoc} to have the same freeze-out temperature in absence of Quintessence $T^{\rm NQ}_{\sss\rm f.o.}\simeq9.35\ {\rm GeV}$. We report in Tab.~\ref{tab:benchmark} the details of the four benchmark models under scrutiny. 
The model with the label {\em bulk} is characterized by a mass splitting with the $A$ Higgs boson $\Delta m_A\equiv(2m_\chi-m_a)/(2m_\chi)\simeq45\%$ and with the NLSP, which is the lightest stau, of $\Delta m_{\tilde\tau_1}\equiv(m_{\tilde\tau_1}-m_\chi)/m_\chi\simeq85\%$, and is therefore outside the resonance or coannihilation regions. 
The model in the {\em stau coannihilation} region achieves nearly complete mass degeneracy between the neutralino and the lightest stau, $\Delta m_{\tilde\tau_1}\simeq0$, while the model in the {\em funnel} region has $\Delta m_A\simeq0$. Finally, the model representative of the {\em higgsino} dark matter case is picked in the representation {\bf 200} and has a higgsino fraction larger than 95\%. 

\begin{table}[b!]
\begin{center}
\begin{tabular}{|c|c|c|c|c|c|c|c|c|}
\hline
& $m_0$ & $M_{1/2}$ & $\tan\beta$ & $\mu$ & $m_\chi$ & $\Omega h^2$ & Wino fract. & Higgsino fract. \\
\hline
Bulk & 500 & 500 & 45.0 & $>0$ & 204 & 0.582 & $<1\%$ & $<0.01\%$\\
Stau Coan. & 327 & 590 & 45.0 & $>0$ & 241 & 0.022 & $<1\%$ & $<0.01\%$\\
Funnel & 408 & 592 & 45.0 & $<0$ & 243 & 0.005 & $<1\%$ & $<0.01\%$\\
Higgsino & 800 & 323 & 45.0 & $>0$ & 250 & 0.009 & $4\%$ & $95.5\%$\\
\hline
\end{tabular}
\end{center}
\caption{The four {\em Benchmark models} representing different possible scenarios for the lightest neutralino. The scalar trilinear coupling $A_0=0$ in all cases, and the neutralino freeze-out temperature in absence of Quintessence is, again for all four scenarios, fixed at $T^{\rm NQ}_{\sss\rm f.o.}\simeq9.35\ {\rm GeV}$.}\label{tab:benchmark}
\end{table}
We analyze in Fig~\ref{fig:casestudyCOMP} the quintessential  relic density enhancement generated in the four benchmark cases as a function of the Quintessence-to-radiation ratio at the neutralino freeze-out temperature without Quintessence ($\xi_\phi$). We let $\xi_\phi$ vary from zero up to $10^5$, highlighting the large and small $\xi_\phi$ regimes, respectively, in the left and right panels. We find that the region of the parameter space where the enhancement is more effective is the coannihilation strip. Between the coannihilation and the bulk regions, we find the intermediate case of the higgsino LSP: this is in fact expected, since coannihilations phenomena take place there as well, though on top of neutralino 
pair-annihilation processes which are by far more efficient than in the bino-like case. Finally, the weakest quintessential enhancement is found in the funnel region.

\FIGURE[t]{\epsfig{file=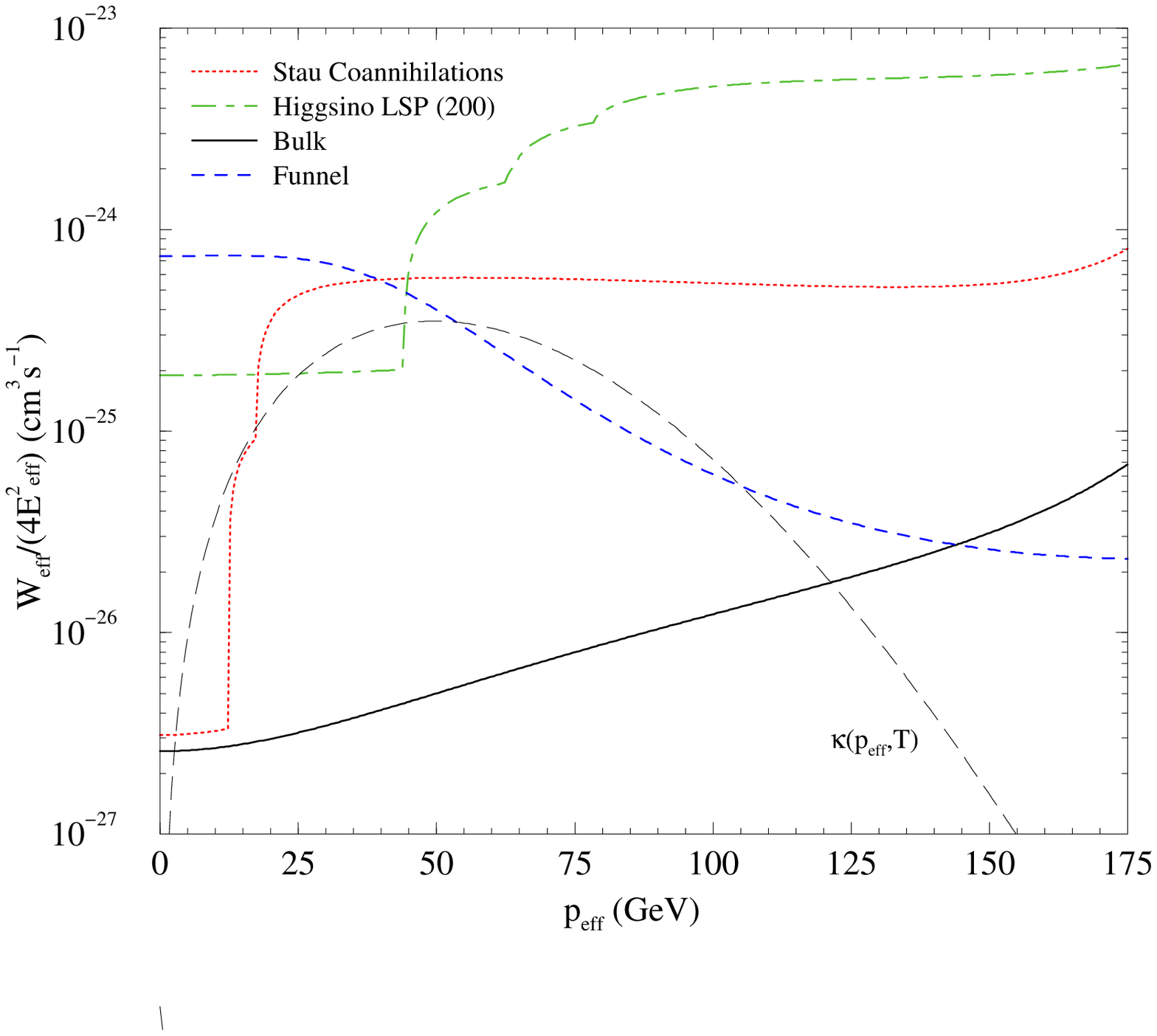,width=7.1cm}\qquad
\epsfig{file=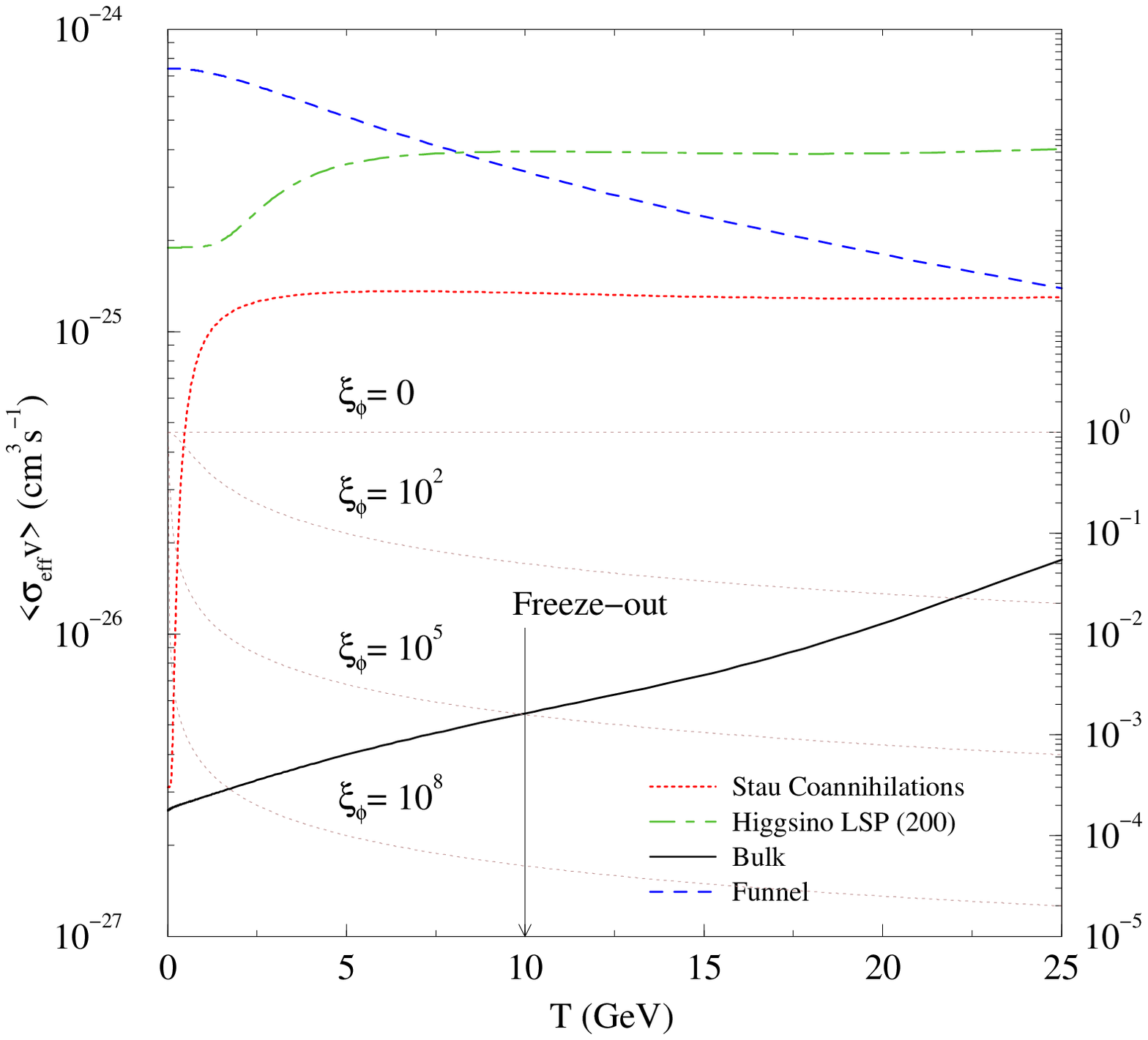,width=7.1cm}
\caption{({\em Left} ): The effective annihilation cross section $W_{\rm eff}/(4E_{\rm eff}^2)$ as a function of $p_{\rm eff}$ for the four benchmark models of Tab.~1. The dashed line indicates the thermal weight factor $\kappa(p_{\rm eff},T)$ at the freeze-out temperature $T^{\rm NQ}_{\rm\sss f.o.}$. ({\em Right}): The thermally averaged effective cross section $\langle \sigma_{\rm eff}v\rangle$ as a function of the temperature for the four benchmark models. The brown dotted lines represent the suppression factor $\left(\sqrt{1+\xi_\phi\frac{T^2}{T^2_{\sss\rm f.o.}}}\right)^{-1})$ due to a quintessential component whose amount at freeze-out is given by $\xi_\phi$; the corresponding reference scale is shown on the right-hand side of the figure.}\label{fig:casestudy}
}


With the purpose of understanding the peculiar pattern emerging from Fig~\ref{fig:casestudyCOMP}, we consider the effective thermally averaged annihilation cross section $\langle\sigma_{\rm eff}v\rangle$, which we rewrite in the following form \cite{Edsjo:2003us}:
\begin{equation}\label{eq:weff}
\langle\sigma_{\rm eff}v\rangle=\int_0^\infty {\rm d}p_{\rm eff}\frac{W_{\rm eff}(p_{\rm eff})}{4 E^2_{\rm eff}}\kappa(p_{\rm eff},T),
\end{equation}  
where $p_{\rm eff}$ and $E_{\rm eff}=\sqrt{p^2_{\rm eff}+m_\chi^2}$ are respectively the three-momentum and the energy of a pair of lightest neutralinos in their center of mass frame. $W_{\rm eff}$ is the effective annihilation rate per unit volume and unit time; in case coannihilation processes are present, it can be written as a weighed sum over all annihilation and coannihilation processes \cite{Edsjo:1997bg}:
\begin{equation}
W_{\rm eff}=\sum_{ij}\frac{4p^2_{ij}}{p_{11}}\frac{g_ig_j}{g_1^2}\sqrt{s}\ \sigma_{ij}.
\end{equation}
Here $g_i$ is the number of internal degrees of freedom for the particle $i$, $p_{ij}$ the common magnitude of the three-momentum in the center of mass frame for the process involving the particles $i$ and $j$, $\sigma_{ij}$ the relative cross section, while the index 1 refers to the LSP (\ie $W_{\rm eff}$ correctly reduces to the annihilation rate $W_{11}$ for a pair of lightest neutralinos, if no coannihilation is present). Finally, eq.~(\ref{eq:weff}) defines the function $\kappa(p_{\rm eff},T)$, which contains the temperature dependence in the Maxwell-Boltzmann approximation. 

We plot in the left panel of Fig~\ref{fig:casestudy} the resulting $W_{\rm eff}/(4E^2_{\rm eff})$ for the four benchmark models of Tab.~\ref{tab:benchmark}. When a coannihilation channel becomes kinematically allowed, the effective cross section abruptly raises, as evident from both the case of the stau coannihilations and that of the higgsino LSP. In this last case, also notice how the cross section at low $p_{\rm eff}$ is larger than in the bulk, or stau coannihilations, case. The figure also shows the typical bell shape of the function $\kappa$, which exponentially suppress the large $p_{\rm eff}$ tails of $W_{\rm eff}$ upon integration over $p_{\rm eff}$. As $T$ is reduced, the peak of the corresponding $\kappa$ function is shifted towards lower $p_{\rm eff}$ values. 
In the right panel of Fig~\ref{fig:casestudy} we plot the result of the integration over $p_{\rm eff}$, obtaining the corresponding curves as functions of $T$ for the various cases. The shape of the resulting curves comes not as a surprise: a decreasing function of $p_{\rm eff}$ (e.g. the case of the funnel) gives rise to a {\em decreasing} function of $T$, and {\em vice-versa}.

\FIGURE[t]{\epsfig{file=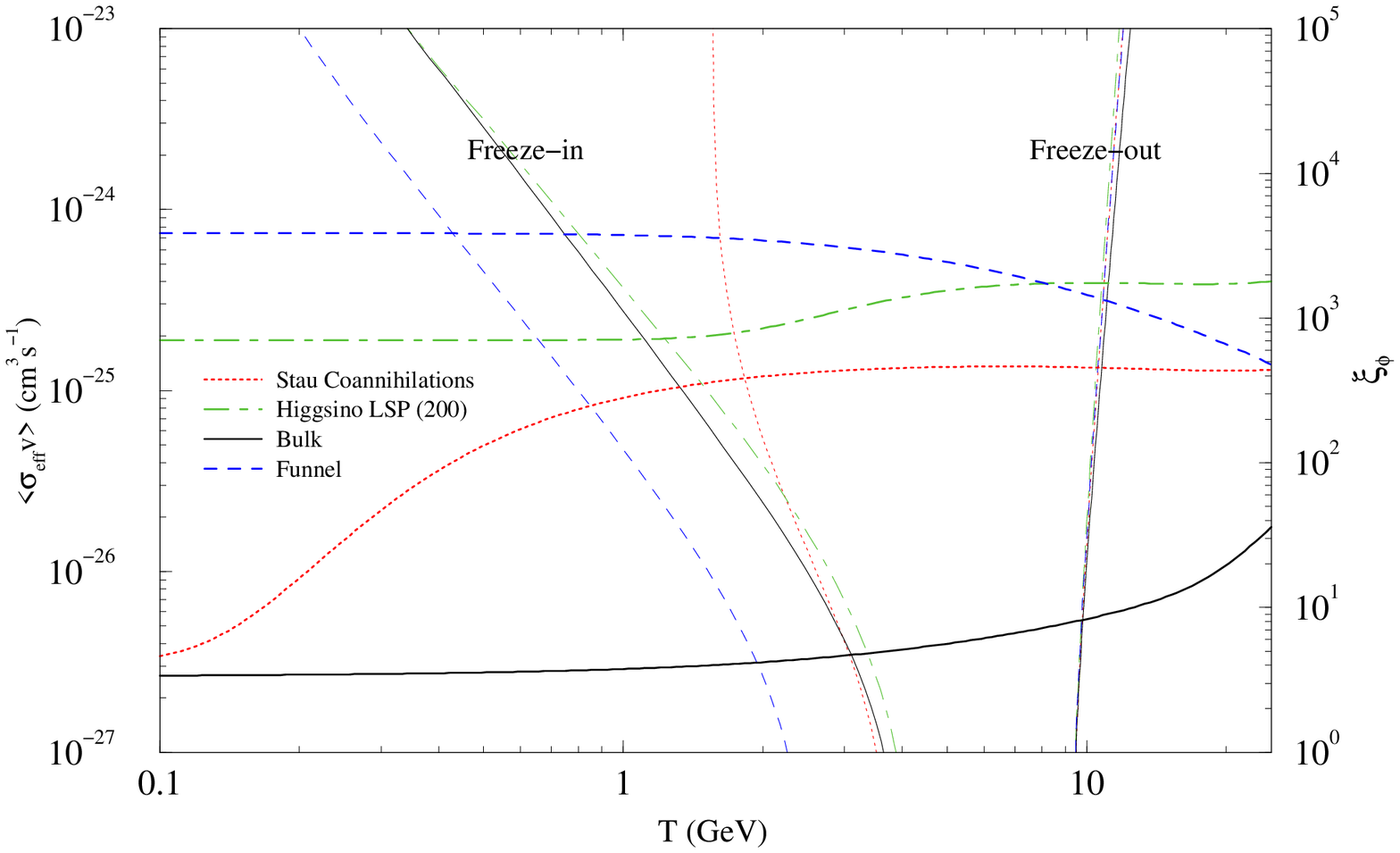,width=11cm}
\caption{A comparison between the thermally averaged cross section $\langle \sigma_{\rm eff}v\rangle$ as a function of the temperature, compared with the temperatures of {\em freeze-out} $T_{\sss\rm f.o.}$ (nearly vertical lines on the right) and of {\em freeze-in} $T_{\sss\rm f.i.}$ (left lying lines). The first one is defined as the temperature when the neutralino number density becomes 50\% larger than the equilibrium number density \cite{Edsjo:2003us}, while $T_{\sss\rm f.i.}$ is the temperature corresponding to the moment when the number density normalized to the entropy density gets 50\% larger than its {\em asymptotic} value, \ie $Y_0$, which gives the nowadays relic density.}\label{fig:wave}
}


The temperature evolution of $\langle \sigma_{\rm eff}v\rangle$ critically enters the thermal history of neutralinos, which, we remind the reader, proceeds from higher to lower temperatures, \ie from right to left in Fig~\ref{fig:casestudy}.
At very large temperatures, the number density of relic species closely follows the equilibrium number density. Afterwards two steps can be identified. First, the number density $n$ starts to become {\em larger} than the equilibrium number density $n_{\rm eq}$, in a  stage commonly called {\em freeze-out}. This step takes place when the quantity $\Gamma\equiv n\langle \sigma_{\rm eff}v\rangle$ becomes of the same order of the expansion rate of the universe $H(T)$. (We recall that we did not define the freeze-out temperature $T_{\sss\rm f.o.}$ from the relation $\Gamma=H$, but rather as the temperature when the neutralino number density becomes 50\% larger than the equilibrium number density, \ie $n=1.5 n_{\rm eq}$ \cite{Edsjo:2003us}). Then, a second step goes on, where (co-)annihilations of neutralinos further reduce the number-to-entropy density $Y$ towards its asymptotic value $Y_{0}$, with an efficiency which depends on the relevant scattering cross sections at temperatures below freeze-out. For definiteness, we label as {\em freeze-in} temperature $T_{\sss\rm f.i.}$ the temperature at which $Y=1.5\ Y_0$.

In our discussion we have so far considered just the first step, \ie the increase in $T_{\sss\rm f.o.}$ set by the shift $H^{\rm NQ}\rightarrow H^{\rm NQ}\sqrt{1+\rho_\phi/\rho_{r}}$. To explain the pattern emerging from Fig~\ref{fig:casestudy}, we need however to consider also the second effect mentioned above. The evolution of $Y$ after the freeze-out temperature is described by the Boltzmann equation (\ref{eq:Boltzmann3}) with $Y_{\rm eq}=0$. Once exactly integrated, one obtains:
\begin{equation}\label{eq:evolutionpostfreezeout}
\frac{1}{Y_0}=\frac{1}{Y_{\sss\rm f.o.}}+\int_{T_0}^{T_{\sss\rm f.o.}}{\rm d}T\frac{\langle \sigma_{\rm eff}v\rangle}{H}\frac{1}{3}\frac{ds}{dT}.
\end{equation}
In presence of Quintessence, the integrand in the right-hand side gets the suppression factor, which similarly to step 1 has the form   
\begin{equation}\label{eq:suppressionfactor}
\frac{1}{H^{\rm NQ}(T)}\rightarrow \frac{1}{H^{\rm NQ}(T)}\frac{1}{\sqrt{1+\rho_\phi/\rho_{r}(T)}}\simeq\frac{1}{H^{\rm NQ}(T)}\frac{1}{\sqrt{1+\xi_\phi T^2/T^2_{\sss\rm f.o.}}}.
\end{equation}

Clearly, the longer the post-freeze-out annihilations last, the larger the suppression of the relic abundance is. The convolution between $\langle \sigma_{\rm eff}v\rangle$ and the suppression factor of eq.~(\ref{eq:suppressionfactor}) sets the shift in the temperature range where neutralino annihilations are effective, down to the {\em freeze-in} temperature. The four benchmark models show very different low temperatures behavior of $\langle \sigma_{\rm eff}v\rangle$: in the case of the funnel, $\langle \sigma_{\rm eff}v\rangle$ {\em increases} at lower temperatures, hence we expect a sharp shift in $T_{\rm\sss f.i.}$ to lower values as $\xi_\phi$ increases; on the other hand, in the case of stau coannihilations, $\langle \sigma_{\rm eff}v\rangle$ rapidly drops to low values at low temperatures, therefore  freeze-in takes place earlier, at higher temperatures. The higgsino LSP and the bulk cases are intermediate, and quite similar to each other. The emerging picture is clarified in Fig~\ref{fig:wave}, which summarizes the two effects we discussed: we depict both the freeze-out and the freeze-in temperatures as functions of $\xi_\phi$, which is shown on the vertical scale at the right-hand side of the plot, as well as the thermally averaged cross sections in the four cases under inspection. While the (slight) hierarchy in the freeze-out temperature depends on the cross sections at temperatures larger than freeze-out, the above mentioned convolution dictates the shape of the freeze-in curves. The pattern emerging from the superposition of the two mentioned effects is the one we found in Fig~\ref{fig:casestudyCOMP}.

Since both the freeze-in and freeze-out temperatures depend on the same factor, which 
 scales with $\xi_\phi$ as $\approx \sqrt{1+\xi_\phi T^2/T^2_{\sss\rm f.o.}}$, we expect to be able to give a rough rule-of-thumb estimate of the quintessential enhancement effects for a given $\xi_\phi$. We do not, however, expect a rigorous scaling as the two intervening effects act on different temperature ranges. The approximate scaling we propose tries to keep track of both the correct limit as $\xi_\phi\rightarrow0$ and the would-be behavior at large $\xi_\phi$:
\begin{equation}\label{eq:approximation}
\Omega^Q/\Omega\approx(1+a\cdot\xi_\phi)^b,\quad b\simeq 0.5 
\end{equation}
We find that the actual enhancement in the full numerical computation reproduces, at {\em large} $\xi_\phi$, our simple ansatz, although with a less steep increase: a best fit procedure gives in fact $0.4\lesssim b \lesssim 0.5$. Moreover, the interplay between the shifts in the freeze-out and freeze-in temperatures (see Fig~\ref{fig:wave}), non-trivially dictates the value of the $a$ parameter. From the overall fit to the benchmark model data we obtain $0.1\lesssim a\lesssim 0.2$. Though the accuracy of resorting to such an approximate formula is necessarily low, eq.~(\ref{eq:approximation}) gives the correct order of magnitude of the quintessential enhancement, and may be of relevance in some cases. For instance, if one deals with a cold dark matter model giving rise to extremely low relic densities, and embeds it in a quintessential cosmological scenario, the formula given in eq.~(\ref{eq:approximation}) may indicate the order-of-magnitude amount of Quintessence at freeze-out which the given model would need in order to get the correct relic density today.

\section{Resurrecting higgsino and wino dark matter}\label{sec:winohiggsino}
When the lightest neutralino is dominated by its higgsino or wino component, the corresponding relic abundance is typically {\em smaller} than
the estimated non-baryonic matter content of the Universe. This is first due to the direct annihilation rate of higgsino-like or wino-like LSPs, which is by far larger than that of a bino-like LSP (see e.g. the left panel of Fig~\ref{fig:casestudy}). In fact, not only the couplings involving, for instance, a wino are larger than those involving a bino, since $g_2>g_1$, but also the number of final states in which winos or higgsinos can annihilate is larger. As an example, winos can annihilate into a couple of weak bosons without any intermediate scalar superpartners, whereas binos cannot. Furthermore, the mass spectrum of a wino-like LSP always yields an approximate mass degeneracy with the lightest chargino, as does a higgsino-like LSP, which is, in its turn, also very close in mass to the next-to-lightest neutralino. This means that coannihilations between the LSP and the lightest chargino, as well as, in the case of the higgsino, with the neutralino $\widetilde\chi_2$, further reduce the relic abundance {\em in the whole parameter space}. This mass degeneracy with the chargino also bears a rather strong lower bound on the neutralino mass through direct accelerator searches. As a result, in the absence of relic density enhancement processes\footnote{Among these, we mention non-thermal neutralino production through direct or indirect decays of gravitinos or moduli fields \cite{etcMurakami:2000me}.}, wino or higgsino LSPs are not adequate to be the main dark matter component. 

In the present framework of quintessential cosmologies, the enhancement resulting from the shift towards larger values of the freeze-out temperature of cold relics may naturally render wino or higgsino LSPs attractive dark matter candidates. In order to elucidate this point, we will discuss two particular cases: mSUGRA with non-universal gaugino masses generated by a gauge kinetic function belonging to the {\bf 200} representation of the symmetric product of two $SU(5)$ adjoints, where the LSP is mostly higgsino-like, and the minimal Anomaly Mediated SUSY Breaking (mAMSB), where the LSP is typically almost completely a wino. We show that (1) the relic abundance naturally falls into the preferred cosmological range for a suitable choice of the quintessential parameters and (2) that the needed {\em fine tuning} in the cosmological parameters is reasonably low, and comparable with that of the supersymmetric parameters.

\subsection{$SU(5)$-like gaugino non-universality}\label{sec:H200}

In the context of supergravity, non-vanishing gaugino masses are generated through the SUSY breaking vev of the auxiliary component of the lowest order non-renormalizable term in the gauge kinetic function. The SUSY breaking vevs, in order to preserve gauge invariance, must lie in a representation of the symmetric product of two adjoints of the underlying unified gauge group. In the case of $SU(5)$ GUTs, one has
\begin{equation}
\left({\bf 24}\times{\bf 24}\right)_{\rm symm}={\bf 1}\oplus{\bf 24} \oplus{\bf 75}\oplus {\bf 200}.
\end{equation}
Universal gaugino masses are generated only if the SUSY breaking field lies in the singlet representation {\bf 1}. In all other cases, particular ratios between gaugino masses will hold at the GUT scale \cite{Ellis:1985jn,Drees:1985bx}. We report in Tab.~\ref{tab:reps} the resulting relative gaugino masses \cite{Anderson:1996bg}.

\begin{table}[b]
\begin{center}
\begin{tabular}{|c|ccc|}
\hline
Rep.&$M_1$&$M_2$&$M_3$\\
\hline
\hline
{\bf 1}&1&1&1\\
{\bf 24}&-1/2&-3/2&1\\
{\bf 75}&-5&3&1\\
{\bf 200}&10&2&1\\
\hline
\end{tabular}
\caption{Relative values of the gaugino masses at the GUT scale for different representations allowed in $SU(5)$.}\label{tab:reps}
\end{center}
\end{table}
Two of the four representations yield a higgsino-like LSP, namely the {\bf 75} and the {\bf 200}. We choose here to focus on the case of the {\bf 200}, because in the case of the {\bf 75} successful electroweak symmetry breaking (EWSB) forces the parameter space to rather narrow regions, and to low values of $\tan\beta$ \cite{Chattopadhyay:2003yk}. 

\FIGURE[t]{\epsfig{file=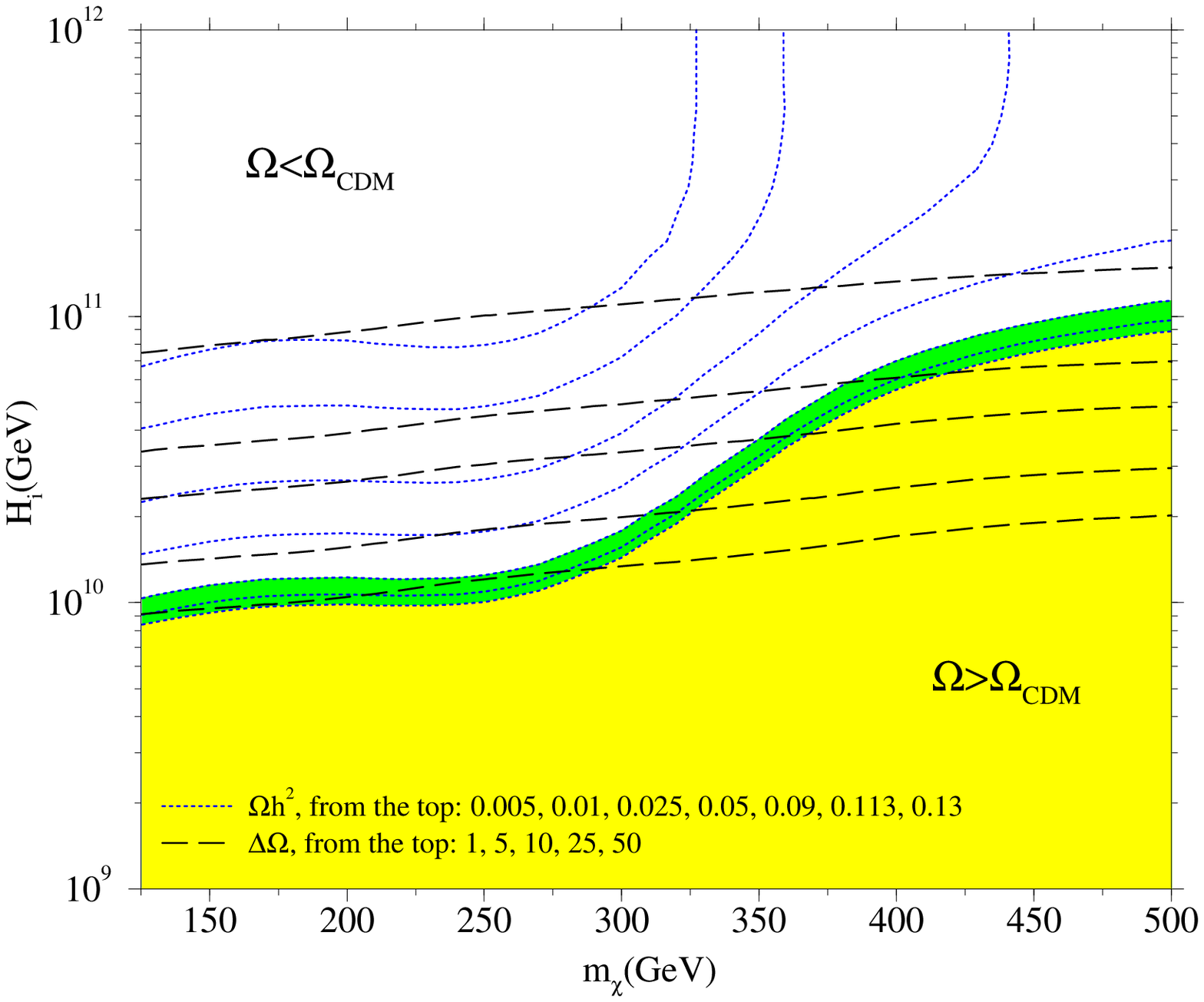,width=7.1cm}\qquad
\epsfig{file=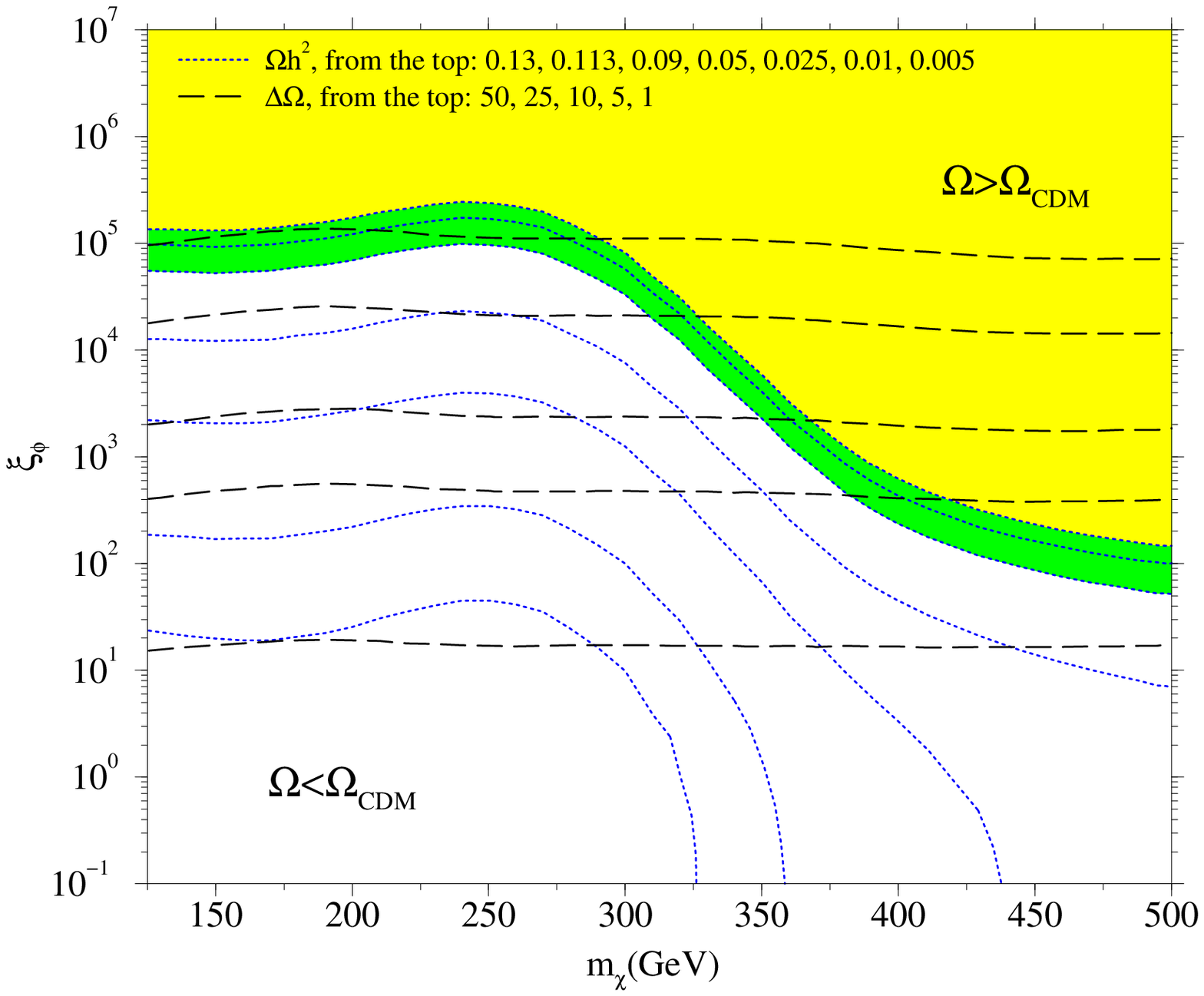,width=7.1cm}
\caption{({\em Left} ): The ($m_\chi,\ H_i$) plane for the case of the representation {\bf 200}, with SUSY parameters $\tan\beta=45$, $A_0=0$, $\mu>0$, $m_0=500\ {\rm GeV}$ and fixing $\lambda=3.5$. The yellow shaded region has $\Omega_\chi h^2>0.13$, while in the green-shaded strip $0.09<\Omega_\chi h^2<0.13$. ({\em Right} ): The ($m_\chi,\ \xi_\phi$) plane at the same values of the SUSY parameters. In both panels, the blue dotted lines correspond to points at fixed $\Omega h^2=0.005,\ 0.01,\ 0.025,\ 0.05,\ 0.09,\ 0.113,\ 0.13$, while the black dashed lines to points at fixed enhancements $\Delta\Omega=1,\ 5,\ 10,\ 25,\ 50$.}\label{fig:H200}
}

We study the requirements on the cosmological sector needed to achieve a neutralino relic density compatible with the current estimates of the dark matter content of the Universe. In particular, we consider a sample case in which we fix 
$\tan\beta=45$, $A_0=0$, $\mu>0$ and $m_0=500\ {\rm GeV}$, and let $M_{1/2}$ vary within the range allowed by EWSB. Changing the value of $m_0$ does not lead to remarkable changes in the results: we only find that at lower values $m_0<500$ GeV and at low $M_{1/2}$ the $A$-funnel condition $m_\chi\simeq m_A/2$ approximately holds, while at larger values of $m_0$ the allowed $M_{1/2}$ range is more and more reduced by the requirement of successful EWSB. In order to have a wider range of neutralino masses, we resort to a large value of $\tan\beta$.

We report our results in Fig~\ref{fig:H200}, where rather than $M_{1/2}$ we consider the neutralino mass $m_\chi$ as the last SUSY parameter which fully fixes our model setup: the mass range displayed is dictated by the fulfillment of phenomenological bounds and of successful radiative EWSB. 
The Quintessence contribution to the expansion rate of the Universe is defined by fixing the parameter $\lambda = 3.5$ in the exponential potential eq.~(\ref{eq:exppot}) and by varying the initial condition on $H_i$.
In the left panel we show the parameter $H_i$, while in the right panel we plot a more model independent result by trading $H_i$ for the Quintessence-to-radiation ratio at neutralino freeze-out $\xi_\phi$. The region shaded in yellow bears an excessive relic density enhancement, giving rise to $\Omega_\chi h^2>0.13$, while the green shaded strip has $0.09<\Omega_\chi h^2<0.13$, and thus approximately reproduces the 2-$\sigma$ range for $\Omega_{\rm\sss CDM}h^2$. In the figures we also plot some isolevel curves both for $\Omega h^2$ (blue dotted lines) and for $\Delta\Omega$ (black dashed lines). At large $H_i$, \ie\ when Quintessence is dominant only at very early times, the iso-$\Omega$ curves tend to become vertical lines, in the region where $\Delta\Omega\ll1$. As expected, we find that the quintessential enhancement, quantified by $\Delta\Omega$, is, to a good approximation, fixed by $\xi_\phi$, as emerging from the $(m_\chi,\xi_\phi)$ plane, where the isolevel curves for $\Delta\Omega$ are practically horizontal. The peculiar behavior of the iso-$\Omega$ curves as functions of $m_\chi$ is due to the fact that, raising $m_\chi$, the higgsino fraction of the lightest neutralino grows, thus partly compensating (at $m_\chi\lesssim250 \ {\rm GeV}$) the effect of a larger mass on the relic density. When the neutralino starts to be largely higgsino-dominated ($m_\chi\gtrsim 250 \ {\rm GeV}$), $\Omega h^2$ monotonously grows with the neutralino mass.
We highlight that at the left sides of the figures, \ie\ at low $m_\chi$, the required enhancement is $\Delta\Omega\approx50$, which means that the neutralino relic density in the absence of Quintessence is two orders of magnitude below $\Omega_{\rm\sss CDM}$. Actually, in the absence of Quintessence, the neutralino relic density, at the smallest value for $m_\chi$ is found to be $\Omega h^2\simeq0.002$. We checked that the {\bf 200} model yields, at any $\tan\beta$, a neutralino relic density lying at least a factor 2 below $\Omega_{\rm\sss CDM}$, and in the specific case $\tan\beta=45.0$, $\Omega_\chi h^2<0.03$: therefore, in this context, a relic density enhancement mechanism is mandatory.  

\subsection{mAMSB and wino LSP}\label{sec:mAMSB}

In the minimal anomaly mediated supersymmetry breaking (mAMSB) model, the soft SUSY breaking contributions arising from the super-Weyl anomaly are supposed to be the dominant ones \cite{Randall:1998uk,Giudice:1998xp}. As a consequence, the gaugino masses are entirely determined by the corresponding $\beta$ functions and by the gravitino mass $m_{3/2}$. The problem of negative scalar masses is cured by adding a universal contribution $m^2_0$. The model is then defined by a set of four parameters, namely
\begin{equation}
m_0,\ m_{3/2},\ \tan\beta\ {\rm and}\ {\rm sign}\mu. 
\end{equation}
Since at the weak scale $M_1/M_2\sim3.2$, the LSP of mAMSB models is dominantly a {\em wino}, with a very large typical purity. The lightest chargino is thus quasi degenerate with the LSP, and the total cross section of neutralino pair annihilation is very large \cite{Gherghetta:1999sw}. The pattern for the quantity $W_{\rm eff}$ is therefore rather similar to that of the higgsino LSP case, a part from the absence of the coannihilations with the next-to-lightest neutralino, and therefore of the second bump visible in the left panel of Fig~\ref{fig:casestudy}. As far as the quintessential enhancement is concerned, we find, as expected, that the enhancement pattern for wino-like neutralinos is very close to that in the higgsino case.

\FIGURE[t]{\epsfig{file=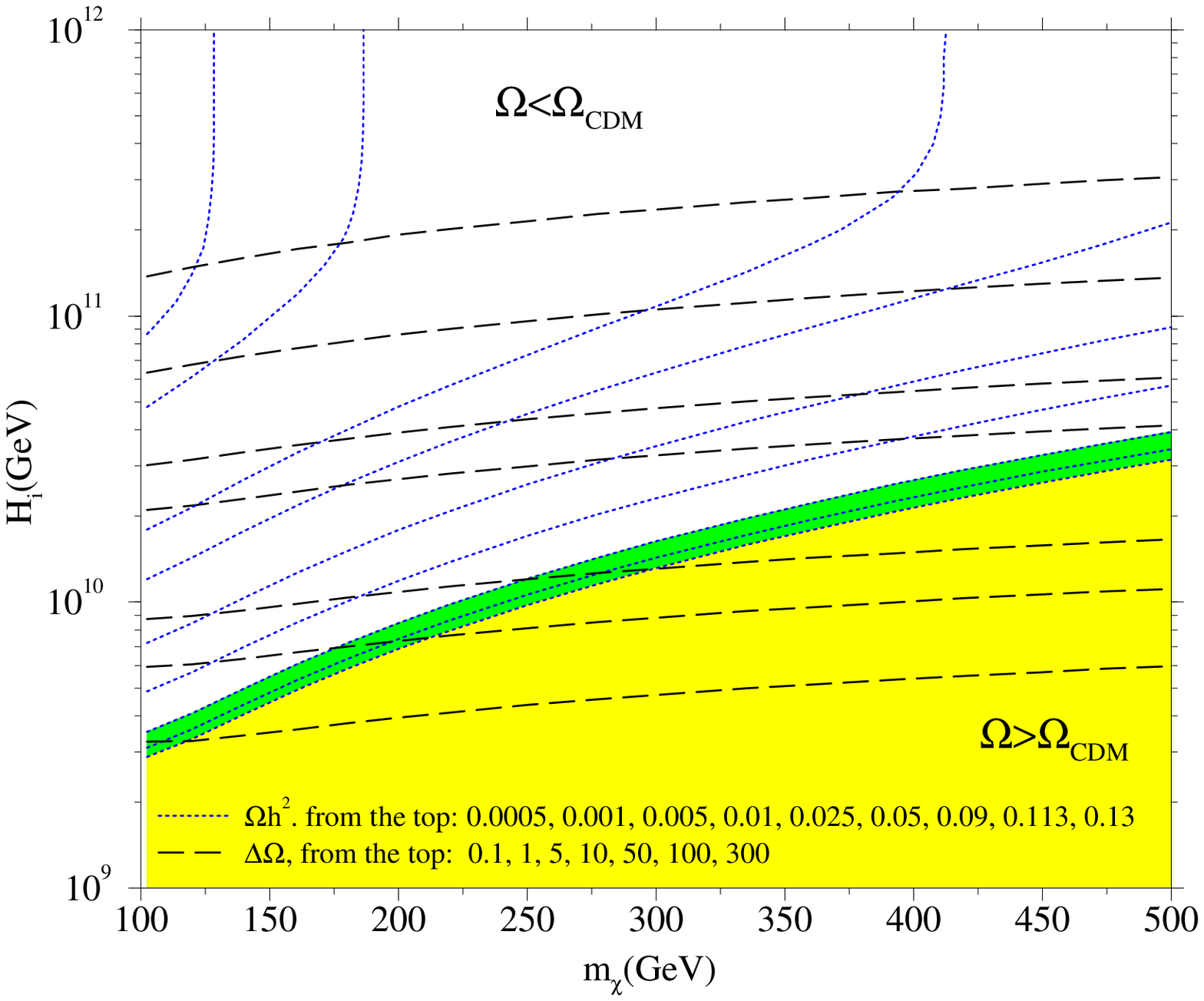,width=7.1cm}\qquad
\epsfig{file=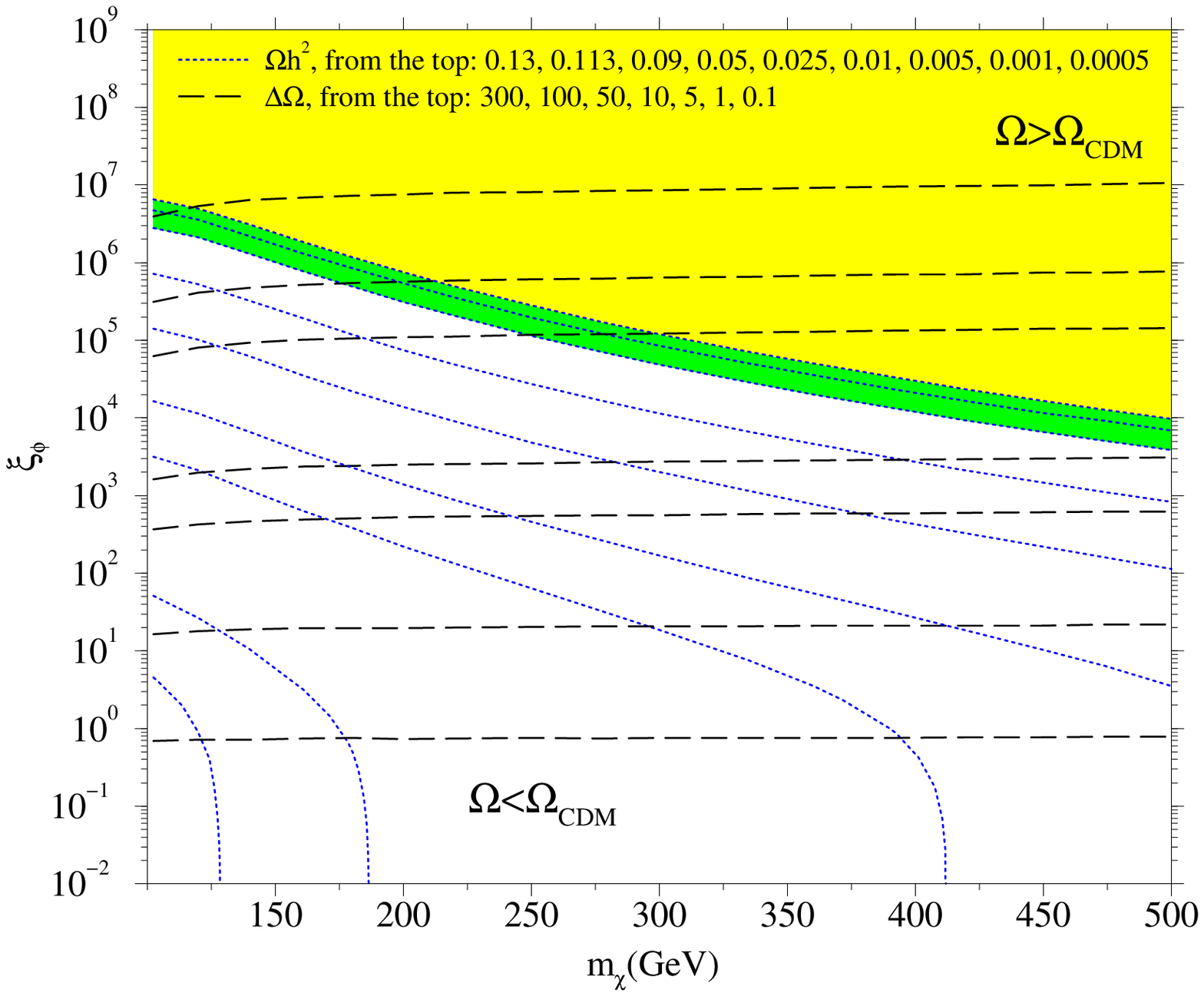,width=7.1cm}
\caption{({\em Left} ): The ($m_\chi,\ H_i$) plane for the case of the anomaly mediated SUSY breaking, at $\tan\beta=50$, $\mu>0$ and $m_0=1\ {\rm TeV}$. The yellow shaded region has $\Omega_\chi h^2>0.13$, while in the green-shaded strip $0.09<\Omega_\chi h^2<0.13$. ({\em Right} ): The ($m_\chi,\ \xi_\phi$) plane at the same values of the SUSY parameters. In both panels, the blue dotted lines correspond to points at fixed $\Omega h^2=0.0005,\ 0.001,\ 0.005,\ 0.01,\ 0.025,\ 0.05,\ 0.09,\ 0.113,\ 0.13$, while the black dashed lines to points at fixed enhancements $\Delta\Omega=0.1,\ 1,\ 5,\ 10,\ 25,\ 50$.}\label{fig:AMSB}
}


We study in Fig~\ref{fig:AMSB} the quintessential relic density enhancement needed to obtain cosmologically viable neutralino relic densities in the wino LSP mAMSB scenario. We focus on a sample case at $m_0=1$~TeV, 
$\tan\beta=5$, $\mu>0$ and $m_{3/2}$ ranging from 32 to 192 TeV. We plot in the figures, as in the previous higgsino case, the neutralino mass range allowed by accelerator searches (lower bound) and by successful EWSB (upper bound). Again we shade the cosmologically preferred region in green, while the yellow shaded parts yield an overproduction of neutralinos. We plot the $(m_\chi,\ H_i)$ plane in the left part of the figure, while $(m_\chi,\ \xi_\phi)$ in the right panel. Also shown are the curves corresponding to particular values of the resulting relic density (blue dotted lines) and of the quintessential enhancement (black dashed lines).  As emerging from the comparison between Fig~\ref{fig:H200} and Fig~\ref{fig:AMSB}, in the present mAMSB case, the needed enhancement factors are {\em larger} than in the higgsino LSP case, since wino pair annihilations have a larger cross section with respect to the higgsinos.
In mAMSB scenarios, the wino-purity does not critically depend on the neutralino mass, thus the iso-$\Omega$ curves are remarkably smooth. The relic density, in its turn, is a growing function of $m_\chi$, and therefore the quintessential enhancement needed to drive the resulting neutralino relic abundance to the cosmologically preferred range  decreases with increasing neutralino masses. We remark that in the present case the neutralino relic density in the absence of Quintessence is always very low, being $\Omega h^2\simeq0.0003$ in correspondence to the lowest neutralino masses. In fact, in the case shown in Fig~\ref{fig:AMSB}, the needed enhancement factor can be as large as $\Delta\Omega\approx 300$.

\subsection{Fine Tuning}

\FIGURE[t]{\epsfig{file=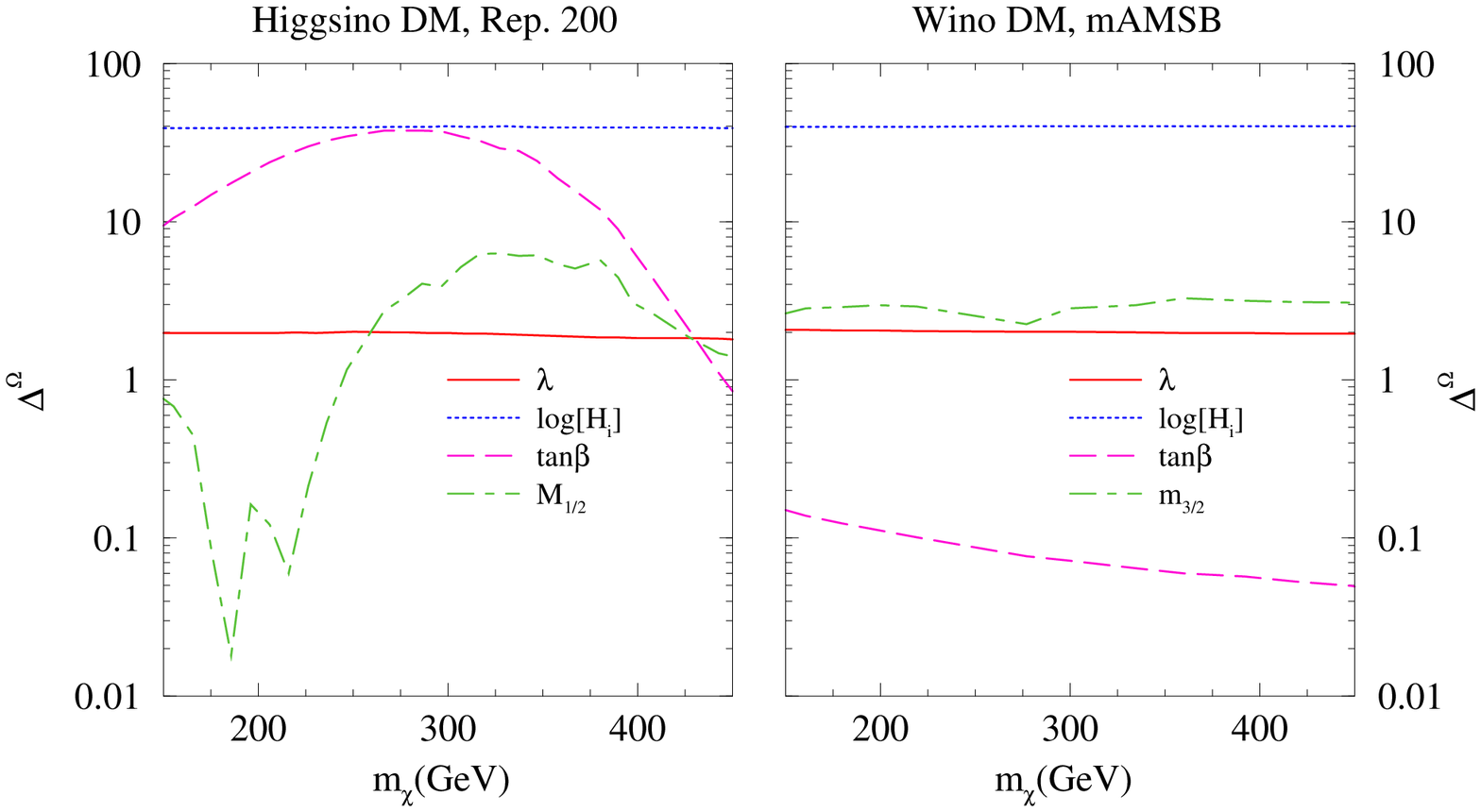,width=9cm}
\caption{The fine tuning $\Delta^\Omega$ with respect to the cosmological parameters $\lambda$ (red solid lines) and $\log[H_i]$ (dotted blue lines), and to the supersymmetric parameter $\tan\beta$ (dashed magenta lines). With dot-dashed green lines, we also show the fine tuning due to $M_{1/2}$ ({\em right}) and to $m_{\sss \frac{3}{2}}$ ({\em left}), respectively ({\em right}) in the case of higgsino dark matter in mSUGRA with non-universal gaugino masses (rep. {\bf 200}) and ({\em left}) in the mAMSB scenario. The amount of fine tuning is computed along the curves at fixed $\Omega h^2=0.113$, as a function of the neutralino mass $m_\chi$.}\label{fig:finetuning}
}


The issue of fine tuning in supersymmetric models of dark matter \cite{Ellis:2001zk} is often quantified through the logarithmic sensitivity of the neutralino relic density with respect to the various parameters appearing in a given model. The amount of fine-tuning for a given phenomenological model is often regarded as a measure of its degree of naturalness. Moreover, the experimental accuracy needed to fix a given parameter is, typically, inversely proportional to the sensitivity of the measured quantity on the parameter itself. It is then of crucial importance to understand whether quintessentially enhanced models of dark matter are ``natural'' or excessively fine-tuned. 
We therefore concentrate, in this section, on the fine tuning in the cosmological sector of a supersymmetric model embedded in a quintessential cosmology. 

To estimate the fine tuning in the parameters defining the cosmological model, \ie\ $H_i$ and $\lambda$, following \cite{Ellis:2001zk} we compute the quantity
\begin{equation}
\displaystyle
\Delta^\Omega\equiv\sqrt{\sum_i\left(\frac{a_i}{\Omega_\chi}\frac{\partial\Omega_\chi}{\partial
a_i}\right)^2}.
\end{equation}
We show our results in Fig.~\ref{fig:finetuning}. The amount of fine-tuning is quantified in the two cases discussed in the preceding sections \ref{sec:mAMSB} and \ref{sec:H200}, in correspondence to the lines at $\Omega h^2=0.113$, \ie\ the central value of the dark matter density as indicated by the latest evaluations \cite{Spergel:2003cb}. The red solid lines refer to the case $a_i=\lambda$, while the blue dotted lines to $a_i=\log H_i$. For comparison, we also include the amount of fine-tuning in the supersymmetric parameters $\tan\beta$ and $M_{1/2}$ ($m_{3/2}$) for the {\bf 200}-higgsino like neutralino (resp. mAMSB). We notice, as expected, that the most fine-tuned parameter is $H_i$, which, in the present context, sets the amount of Quintessence in the early Universe. The dependence on $\lambda$ is, on the other hand, found to be not critical, yielding a moderate fine-tuning in both the higgsino and wino dark matter cases. As compared with the supersymmetric parameters taken into account, we find that the fine-tuning in $H_i$ is of the same order of magnitude of that of $\tan\beta$ in the higgsino-like neutralino case, while in the context of mAMSB the dependence on $\tan\beta$ is largely suppressed: this is traced back to the fact that in the higgsino case, both $M_{1/2}$ and $\tan\beta$ set the higgsino fraction in the lightest neutralino, and therefore, due to the large difference in the pair annihilation of a bino and of a higgsino, they play a crucial r\^ole in determining $\Omega h^2$. The sensitivity of the neutralino relic density to the quintessential parameters is thus found to be of a certain relevance only for the parameter $H_i$, though the corresponding fine-tuning may be of the same order as that needed in the supersymmetric parameters in the sample models considered here or even in the standard mSUGRA setup: in particular even larger amounts of fine tuning are in order in the focus-point, funnel or sfermion coannihilation 
regions, see, e.g.,~\cite{baerfin,Ellis:2001zk}.

\section{Conclusions}

We analyzed the quintessential enhancement of the neutralino relic abundance within ``realistic'' cosmological models. We chose to resort to models of Quintessence exhibiting tracking solutions stemming from an exponential potential, and characterized by only two parameters: the exponent appearing in the potential and the Hubble rate at the end of inflation. In this case, we numerically computed the evolution for the expansion rate parameter and interfaced the result with a fully-numerical computation of the neutralino relic abundance, with an overall estimated accuracy of the order of 1\% or better. We defined a suitable parameter
to quantify the enhancement in the relic abundance induced by Quintessence and
found the limit above which a Quintessence component becomes relevant:
non-negligible effects are expected, at the 1\% or higher, whenever the
ratio in the Quintessence to radiation energy densities at neutralino
decoupling is larger than about 0.1.
Furthermore we showed that, within our viable cosmological models, the enhancement in the neutralino relic density can be as large as six orders of magnitude. Then, we systematically analyzed the dependence of the enhancement on the features of the lightest neutralino as a dark matter candidate (such as its {\em mass} and {\em composition} or any particular mechanism, {\em e.g.} coannihilations or resonance effects, playing a r\^ole in the neutralino relic abundance), concluding that the largest effects take place at large masses and in the coannihilation regions. We also showed that models with a dominantly {\em wino}- or {\em higgsino}-like lightest neutralino (as it the case in mAMSB or in $SU(5)$-inspired gaugino non-universality with the {\bf 75} or the {\bf 200} representations) may well be compatible with the required dark matter content of the Universe if a quintessential enhancement is present. We also showed that, in the proposed models, the fine-tuning in the cosmological parameters is always rather low.

\acknowledgments{We acknowledge fruitful discussions with C.~Baccigalupi.
P.U.\ was supported in part by the RTN project under 
grant HPRN-CT-2000-00152 and by the Italian INFN under the
project ``Fisica Astroparticellare''.}


\begin{thebibliography}{999}


\bibitem{Spergel:2003cb}
D.~N.~Spergel {\it et al.},
{\tt arXiv:astro-ph/0302209}.

\bibitem{sne}
S.~Perlmutter {\it et al.}, Astrophys. J. {\bf 517} (1999) 565;
A.~G.~Riess {\it et al.}, Astron. J. {\bf 116} (1998) 1009.

\bibitem{firstexp}
B. Ratra and P.J.E. Peebles, Phys.\ Rev.\ D\ {\bf 37} (1988) 3406;
C. Wetterich, Nucl. Phys. B {\bf 302} (1988) 668.

\bibitem{firstquint}
P.~J.~E.~Peebles and  B.~Ratra, Astrophys. Jour. {\bf 325} (1988) L17; 
M.~S.~Turner and M.~White, Phys. Rev. D\ {\bf 56} (1997) 4439;
R.~R.~Caldwell, R.~Dave, and P.~J.~Steinhardt, 
Phys. Rev. Lett. {\bf 80} (1998) 1582;
E.~J.~Copeland, A.~R.~Liddle, and D.~Wands, Phys. Rev. D\ {\bf 57}
(1998) 4686.

\bibitem{Steinhardt:nw}
P.~J.~Steinhardt, L.~M.~Wang and I.~Zlatev,
Phys.\ Rev.\ D {\bf 59} (1999) 123504
[arXiv:astro-ph/9812313].

\bibitem{Ferreira:1997hj}
P.~G.~Ferreira and M.~Joyce,
Phys.\ Rev.\ D {\bf 58} (1998) 023503
[arXiv:astro-ph/9711102].


\bibitem{Salati:2002md}
P.~Salati,
[arXiv:astro-ph/0207396].


\bibitem{others}
R.~R.~Caldwell {\it et al.}, Astrophys. J. {\bf 591} (2003) L75
[arXiv:astro-ph/03020505];
L.~Anchordoqui, H.~Goldberg, [arXiv:hep-ph/0306084].

\bibitem{Barrow:ei}
J.~D.~Barrow,
Nucl.\ Phys.\ B {\bf 208} (1982) 501.

\bibitem{Kamionkowski:1990ni}
M.~Kamionkowski and M.~S.~Turner,
Phys.\ Rev.\ D {\bf 42} (1990) 3310.

\bibitem{Gondolo:2000ee}
P.~Gondolo, J.~Edsjo, L.~Bergstrom, P.~Ullio and E.~A.~Baltz,
[arXiv:astro-ph/0012234].

\bibitem{Gondolo:2002tz}
P.~Gondolo, J.~Edsjo, P.~Ullio, L.~Bergstrom, M.~Schelke and E.~A.~Baltz,
[arXiv:astro-ph/0211238].

\bibitem{Edsjo:2003us}
J.~Edsjo, M.~Schelke, P.~Ullio and P.~Gondolo,
JCAP {\bf 0304} (2003) 001
[arXiv:hep-ph/0301106].


\bibitem{kina}
M.~Joyce, Phys. Rev. D\ {\bf 55} (1997) 1875;
M.~Joyce and T.~Prokopec, Phys. Rev. D\ {\bf 57} (1998) 6022.


\bibitem{gondologelmini}
P.~Gondolo and G.~Gelmini, Nucl.\ Phys.\ B\ {\bf 360} (1991) 145.


\bibitem{spokoiny}
B.~Spokoiny, Phys. Lett. {\bf B315} (1993) 40.


\bibitem{infquint}
E.J. Copeland, A.R. Liddle and J.E. Lidsey, (2001)
Phys.\ Rev.\ D\ {\bf 64} (2001) 023509;
G. Huey and J. Lidsey, Phys. Lett. B {\bf 514} (2001) 217;
V. Sahni, M. Sami and T. Souradeep, Phys. Rev. D\ { \bf 65} (2002) 023518.



\bibitem{Albrecht:1999rm}
A.~Albrecht and C.~Skordis,
Phys.\ Rev.\ Lett.\  {\bf 84} (2000) 2076
[arXiv:astro-ph/9908085].

\bibitem{Sahni:1999qe}
V.~Sahni and L.~M.~Wang,
Phys.\ Rev.\ D {\bf 62} (2000) 103517
[arXiv:astro-ph/9910097].

\bibitem{rosati}
F.~Rosati,
Phys.\ Lett.\ B {\bf 570} (2003) 5
[arXiv:hep-ph/0302159].

\bibitem{Peiris:2003ff}
H.~V.~Peiris {\it et al.},
Astrophys.\ J.\ Suppl.\  {\bf 148} (2003) 213
[arXiv:astro-ph/0302225].


\bibitem{Binetruy:1983jf}
P.~Binetruy, G.~Girardi and P.~Salati,
Nucl.\ Phys.\ B {\bf 237} (1984) 285.

\bibitem{Griest:1990kh}
K.~Griest and D.~Seckel,
Phys.\ Rev.\ D {\bf 43} (1991) 3191.

\bibitem{Edsjo:1997bg}
J.~Edsjo and P.~Gondolo,
Phys.\ Rev.\ D {\bf 56} (1997) 1879
[arXiv:hep-ph/9704361].

\bibitem{Ellis:1985jn}
J.~R.~Ellis, K.~Enqvist, D.~V.~Nanopoulos and K.~Tamvakis,
Phys.\ Lett.\ B {\bf 155} (1985) 381.

\bibitem{Drees:1985bx}
M.~Drees,
Phys.\ Lett.\ B {\bf 158} (1985) 409.


\bibitem{etcMurakami:2000me}
B.~Murakami and J.~D.~Wells,
Phys.\ Rev.\ D {\bf 64}, 015001 (2001);  
T.~Moroi and L.~Randall,
Nucl.\ Phys.\ B {\bf 570}, 455 (2000);  
M.~Fujii and K.~Hamaguchi,
Phys.\ Lett.\ B {\bf 525}, 143 (2002);  
M.~Fujii and K.~Hamaguchi,
Phys.\ Rev.\ D {\bf 66}, 083501 (2002);    
R.~Jeannerot, X.~Zhang and R.~H.~Brandenberger,
JHEP {\bf 9912}, 003 (1999);  
W.~B.~Lin, D.~H.~Huang, X.~Zhang and R.~H.~Brandenberger,
Phys.\ Rev.\ Lett.\  {\bf 86}, 954 (2001). 



\bibitem{Anderson:1996bg}
G.~Anderson, C.~H.~Chen, J.~F.~Gunion, J.~Lykken, T.~Moroi and Y.~Yamada,
{\tt arXiv:hep-ph/9609457}.


\bibitem{Chattopadhyay:2003yk}
U.~Chattopadhyay and D.~P.~Roy,
arXiv:hep-ph/0304108.



\bibitem{Randall:1998uk}
L.~Randall and R.~Sundrum,
Nucl.\ Phys.\ B {\bf 557} (1999) 79
[arXiv:hep-th/9810155].

\bibitem{Giudice:1998xp}
G.~F.~Giudice, M.~A.~Luty, H.~Murayama and R.~Rattazzi,
JHEP {\bf 9812} (1998) 027
[arXiv:hep-ph/9810442].

\bibitem{Gherghetta:1999sw}
T.~Gherghetta, G.~F.~Giudice and J.~D.~Wells,
Nucl.\ Phys.\ B {\bf 559} (1999) 27
[arXiv:hep-ph/9904378].


\bibitem{Ellis:2001zk}
J.~R.~Ellis and K.~A.~Olive,
Phys.\ Lett.\ B {\bf 514} (2001) 114
[arXiv:hep-ph/0105004].

\bibitem{baerfin} 
H. Baer, C. Balazs and A. Belyaev, JHEP 0203 (2002) 042


\end{thebibliography}
\end{document}